\documentclass[times,twocolumn,final]{elsarticle}

\usepackage{multirow}
\usepackage{multicol}
\usepackage{amsmath}
\usepackage{mathtools}

\usepackage{setspace}

\usepackage{medima}
\usepackage{framed}

\usepackage[T1]{fontenc}
\usepackage[utf8]{inputenc}

\usepackage{graphicx}
\usepackage[export]{adjustbox}

\usepackage{ragged2e}

\usepackage{amssymb}
\usepackage{pifont}
\usepackage{amssymb}
\usepackage{latexsym}
\usepackage{xurl}
\usepackage[T1]{fontenc}
\usepackage[utf8]{inputenc}

\usepackage{array}
\newcolumntype{H}{>{\setbox0=\hbox\bgroup}c<{\egroup}@{}}

\usepackage{url}
\usepackage{xcolor}

\usepackage{algorithm}
\usepackage{algpseudocode}

\definecolor{newcolor}{rgb}{.8,.349,.1}

\journal{Author Version (preprint)}

\begin{document}

\verso{Bill Cassidy \textit{et~al.}}

\begin{frontmatter}

\title{Gaussian Random Fields as an Abstract Representation of Patient Metadata for Multimodal Medical Image Segmentation}

\author[1]{Bill \snm{Cassidy}\corref{cor1}} 
\cortext[cor1]{Corresponding author:}
\ead{bill.cassidy@stu.mmu.ac.uk}


\author[1]{Christian \snm{McBride}}
\author[1]{Connah \snm{Kendrick}}
\author[2]{Neil D. \snm{Reeves}}
\author[3]{Joseph M. \snm{Pappachan}}
\author[1]{Shaghayegh \snm{Raad}}
\author[1]{Moi Hoon \snm{Yap}}

\address[1]{Department of Computing and Mathematics, Manchester Metropolitan University, John Dalton Building, Chester Street, Manchester, M1 5GD, UK}
\address[2]{Medical School, Faculty of Health and Medicine, Health Innovation Campus, Lancaster University, LA1 4YW, UK}
\address[3]{Lancashire Teaching Hospitals NHS Foundation Trust, Preston, PR2 9HT, UK}

\received{XX XXX 2025}
\finalform{XX XXX 20XX}
\accepted{XX XXX 20XX}
\availableonline{XX XXX 20XX}

\begin{abstract}
The growing rate of chronic wound occurrence, especially in patients with diabetes, has become a concerning trend in recent years. Chronic wounds are difficult and costly to treat, and have become a serious burden on health care systems worldwide. Chronic wounds can have devastating consequences for the patient, with infection often leading to reduced quality of life and increased mortality risk. Innovative deep learning methods for the detection and monitoring of such wounds have the potential to reduce the impact to both patient and clinician. 
We present a novel multimodal segmentation method which allows for the introduction of patient metadata into the training workflow whereby the patient data are expressed as Gaussian random fields. 
Our results indicate that the proposed method improved performance when utilising multiple models, each trained on different metadata categories. Using the Diabetic Foot Ulcer Challenge 2022 test set, when compared to the baseline results (intersection over union = 0.4670, Dice similarity coefficient = 0.5908) we demonstrate improvements of +0.0220 and +0.0229 for intersection over union and Dice similarity coefficient respectively. 
This paper presents the first study to focus on integrating patient data into a chronic wound segmentation workflow. Our results show significant performance gains when training individual models using specific metadata categories, followed by average merging of prediction masks using distance transforms. All source code for this study is available at: https://github.com/mmu-dermatology-research/multimodal-grf 
\end{abstract}


\end{frontmatter}


Chronic wounds are a serious condition that can expose patients to infection and potentially increased mortality risk \cite{petersen2022mortality}. The global diabetes epidemic is an important factor in the case of chronic wounds, as patients with diabetes are both at increased risk of developing such wounds and are likely to experience significantly impaired healing rates \cite{dasari2021wound}. 


Patients who have been diagnosed with diabetic foot ulcers (DFU) have been shown to have significantly greater mortality risk when compared to those without \cite{chammas2016mortality}. Such patients are also more likely to suffer from additional comorbidities such as cardiovascular disease, peripheral arterial disease, retinopathy, and neuropathy \cite{brownrigg2013pad, su2018dfu, serban2020retinopahy, chin2024mortality, fan2024risk}. 

Arterial leg ulcers (ALU), DFU, and venous leg ulcers (VLU) can lead to impaired quality of life \cite{franks2016venous, moura2019dfu}. Occurrence of such wounds is associated with an incidence increase of amputation and subsequent mortality risk. These factors are particularly prevalent in older patients, and those suffering from anemia and peripheral artery disease \cite{franks2016venous, costa2017mortality, vainieri2020mortality}. 
Prevalence of chronic wounds is linked to increased occurrence of emotional and physical burdens on patients \cite{renner2017depression, polikandrioti2020depression}. Depression is also commonly associated with patients with chronic wounds \cite{iversen2015depression, iversen2020telemedicine}. 

Chronic wounds are often associated with comorbidities such as vascular deficits, hypertension, chronic kidney disease, and diabetes \cite{sen2021wound}. Diabetic neuropathy is often present in patients with DFU and is a primary cause of DFU \cite{petrone2021foot}, meaning that patients lose sensation in their feet resulting from nerve damage \cite{rathur2007neuropathic}. Patients with DFU may therefore go through prolonged periods without realising that the DFU is present until the wound worsens, frequently leading to other major complications. Wound infection occurs in more than 50\% of all DFU cases \cite{bader2008diabetic} and is one of the most common causes of hospitalisation for  patients with diabetes \cite{petrone2021foot}. Diabetic leg and foot wounds are amongst the most expensive types of wounds to treat in the United States \cite{sen2021wound}, while venous leg ulcers exhibit a recurrence rate as high as 70\% \cite{franks2016venous} within 3 months after wound closure. 

Chronic wound management can be challenging, especially for wounds that have not been identified at an early stage. Such wounds require more intensive treatment programmes, resulting in increased treatment costs. Cases of this type require frequent visits to hospitals or clinics where wounds are assessed by clinical wound experts \cite{boulton2005burden, netten2017mobile}. 
Once wounds have fully healed, recurrence rates are high and can often lead to serious infection and amputation \cite{apelqvist1993prognosis, larsson1998prognosis}. 
In an effort to meet these increased demands on clinics and hospitals, there has been a growing research interest concerning non-contact automated deep learning detection and monitoring of chronic wounds \cite{reeves2021diabetes}. 

The utilisation of deep learning methods to provide a means of early detection and remote wound monitoring could be a gateway to help reduce risks to patients who are vulnerable, and to ease the burdens that clinics and hospitals are currently experiencing \cite{yammine2021telemedicine}. Low cost consumer mobile devices can be used to bring such technologies to patients living in poorer regions, where access to expert healthcare services may be limited. Such advances could also be used to promote patient engagement, a facet of patient care that has been shown to be an effective treatment strategy \cite{marzban2022patient}. 



\section{Related Work}
\label{sec2}
Chronic wounds deep learning research has become increasingly prominent over the last decade \cite{joseph2022future}. Researchers have made advances in tasks such as classification \cite{cassidy2021eval}, localisation 
\cite{yap2021evaluation}, and segmentation \cite{cassidy2023summary, brungel2025overview}. 
More recently, researchers have focussed on integrating patient metadata into convolutional neural network (CNN) training workflows in numerous medical imaging domains. In this section, we explore a selection of the most relevant studies. 


\subsection{Metadata}

Gessert et al. \cite{gessert2020efficientnets} achieved first place in the ISIC 2019 Skin Lesion Classification Challenge \cite{codella2019isic} which required the use of patient clinical details as part of the training workflow. Their classification method, used for task 2 in the challenge, utilised both dermoscopic skin lesion images and additional associated patient meta data which they incorporated using a 2-layer dense neural network (DNN) branch. 
To ensure that the DNN did not associate missingness with the ``unknown'' lesion class (which had no associated patient metadata), they randomly encoded individual properties as missing using a probability of $p = 0.1$. Their validation results indicated that one-hot encoding of age groups gave worse performance metrics. The CNN trained using the dermoscopy images was frozen, then the meta-data DNN output was concatenated with the CNN feature vector prior to dense and classification layers. Finally, the attached DNN and the classification layer was trained. A weighted cross-entropy loss function was used to address class bias where underrepresented classes were assigned a greater weight-based frequency during training. When comparing their CNN results with the CNN + DNN results, and taking into account all classes, the test results showed a reduced performance in mean AUC (-0.0194), and mean sensitivity (-0.0547). However, a performance increase was measured for specificity (+0.0059). For individual classes, the highest performance increases in AUC were observed for melanoma, actinic keratosis, benign keratosis, and dermatofibroma. No performance improvements were measured for individual classes in terms of sensitivity, but all classes (except the ``unknown'' class) saw improved performance. 


Ha et al. \cite{ha2020identifying} utilised an ensemble of highly diverse CNN and non-CNN meta-data models to win the SIIM-ISIC Melanoma Classification Challenge 2020. For this binary classification task (benign vs malignant), they trained their model for 9-class multi-class classification, and used the melanoma class probability to determine the final binary classification results. To boost model performance, they introduced 14 meta-data features into some of the sub-models. Features included sex, age, and anatomical site of the lesion. They also included meta-data related to the images, such as image size (in bytes) and number of images present in the dataset that pertain to the patient each image relates to. 
The ensemble approach averaged the model probability ranks where the probability predictions from each model were uniformly converted to a range of [0,1] prior to averaging. This study observed that models trained using only images performed better than the metadata models overall. However, the inclusion of metadata models in the ensemble experiments provided good model diversity and boosted the overall performance with an AUC of 0.9600. 

Lemay et al. \cite{lemay2021metadata} adapted a feature-wise linear modulation conditioning method for medical image segmentation enabling the integration of metadata into U-Net spinal cord tumour segmentation models. The metadata is used to modulate the segmentation process using low-cost affine transformations which are then applied to feature maps during training which can be used in any CNN architecture. They found that the application of a single metadata item (tumour type) as an additional input into the segmentation network provided a 5.1\% boost to performance. 

Ningrum et al. \cite{ningrum2021metadata} developed a malignant melanoma binary detection model trained using dermoscopic images and patient meta-data targeted at low-resource devices. Using 1200 ISIC skin lesion images and corresponding patient meta-data, they demonstrated an accuracy increase of 18.65\% when comparing a CNN model with a CNN model concatenated with assisting meta-data features from a standard artificial neural network (ANN). They found the concatenation of the outputs of the two different models helped to reduce overfitting when compared to the CNN results when trained only on images. 

Anisuzzaman et al. \cite{anisuzzaman2022multimodal} used wound location data to improve performance of a multi-class wound classification model using two publicly available chronic wound datasets (AZH and Medetec) and a private wound location meta-data dataset. 
For the corresponding categorical body map meta-data, each wound location was converted to an integer value and encoded using one-hot encoding. 
Two main models were trained, a CNN for the wound images, and a second for the wound location meta-data. The CNN utilised transfer learning with VGG-16, VGG-19, ResNet-50 and IncpetionV3 sub-models together with an AlexNet. The wound location model utilised a Multi-Layer Perceptron (MLP) and a Long Short-Term Memory (LSTM) model. The output of the pretrained CNNs was concatenated with the outputs of AlexNet, the MLP, and the LSTM to form the final predictions. Their experiments demonstrated an improvement in classification performance from 72.95\% to 97.12\% when body map meta-data features were introduced into the training workflow. They also completed experiments using the one-hot vector as a direct input into the CNN dense layer. However, this resulted in inferior results when compared to utilising the MLP or the LSTM. 

Patel et al. \cite{patel2024wound} would later build on the prior work completed by \cite{anisuzzaman2022multimodal}. They proposed an improved multi-class multi-modal network architecture utilising parallel spatial and channel-wise squeeze and excitation, axial attention, and an adaptive gated multi-layer perceptron. These modifications allow the network to capture global contextual information by focusing on channel interdependancies, learning patterns across different input channels. 
Spatial information is also maintained by focusing on the spatial
interdependancies of individual channels. Using the AZH and Medetec datasets they achieved an accuracy of 73.98-100\% for classification using assisting location meta-data. However, as per the previous study conducted by \cite{anisuzzaman2022multimodal}, the dataset used was relatively small, with just 930 AZH and 538 Medetec images. 

Gu et al. \cite{gu2024multi} proposed a multimodal architecture capable of simultaneous segmentation and diagnosis of tumours using images and patient metadata. Their architecture comprised of three parts: an image encoder, a text encoder, and a decoder utilising an interlaced sparse self-attention mechanism. Text embedding features are fused at several network layers with the image data between the two encoders. The text preprocessing block embeds the metadata using a language model with bidirectional encoder representations from transforms (BERT). Each word is converted into a two-dimensional vector, with a $2\times$ upsample applied using deconvolution so that the size matches that of the input images. They reported significant improvements for basal cell carcinoma segmentation on two private datasets: +14.3\% IoU and +8.5\% DSC on the ZJU2 dataset, and +7.1\% IoU on the LIZHU1 dataset. They also demonstrated state of the art performance on the GlaS dataset for gland segmentation in colon histology images (DSC +0.0018, IoU +0.0027). A major limitation of this work, however, is the limited size of the datasets used, with each dataset comprising fewer than 200 images. 

Research of this nature demonstrates the potential of the inclusion of metadata in the development of multimodal CNNs. However, as indicated by previous research, there is a severe lack of publicly available multimodal datasets, an issue which is even more apparent in the case of chronic wounds. 

\subsection{Random Fields}
Random fields are a generalised form of a stochastic field where randomness is determined as a function of spatial variables. Essentially, they encompass a random variation of measurable properties \cite{xi2012random}. Gaussian random fields (GRF) are a type of random field that provide a statistical tool to describe naturally occurring structures exhibiting spatially varying uncertainties \cite{latz2019gaussian}. 
They represent a description of uncertainties that can exert critical impacts on the overall performance of physical processes found throughout nature \cite{liu2019gaussian}. 
GRFs are used to model uncertainties such as material properties, measurement errors, and distributions of attributes associated with living organisms. 
Treating such uncertainties as random fields or random variables, statistical analyses can be utilised more consistently \cite{stefanou2009stochastic, tang2015slope}. Practical applications of GRFs include modelling of landscapes in ecology and generation of cloud features in geoscience \cite{li2012cloud}. Variants of GRFs, known as Euclidean bosonic massless free fields, are used for modelling random surfaces and functions in quantum field theory \cite{scott2004fields}. 

GRFs have previously been used in active learning and semi-supervised training processes. Zhu et al. \cite{zhu2003gaussian} proposed a GRF text and digit classification model defined with respect to a weighted graph which represented labelled and unlabelled data. Their framework exploited unlabelled data structures to enhance classification accuracy. However, these experiments predate more modern deep learning methods. 

In more recent works, Yang et al. \cite{yang2023markov} conducted classification experiments on synthetic aperture radar images and improved performance using Markov random fields (MRF). MRF is a random field variant of GRF that introduces a Markov property which models a prediction based only on the current state and not prior or later states. 
This work proposed the generation of a probability field which describes regional relationships. 
They derived the energy function using the intensity field and the probability field, allowing for superior initialisation of the MRF. 

GRFs have been used in multiple scientific disciplines, however, to the best of our knowledge they have not been used to represent metadata in multimodal deep learning experiments. 


\section{Method}
In this section we detail the training, validation, and testing workflow, the proposed method for generating GRFs using patient metadata, and corresponding metrics used to assess our multimodal chronic wound segmentation experiments.

\subsection{Patient Metadata}
Our experiments utilise the following patient metadata categories: (1) patient date of birth (DOB); (2) patient gender (male or female); (3) Health and Disability Decile (HDD). HDD values were obtained from the English indices of deprivation 2019 public records \cite{2019deprivation} using patient post codes as reference. All patient metadata are present for all associated chronic wound images, with no missing instances.

\subsection{Data Normalisation}
\begin{figure}[!h]
  \centering
  \begin{tabular}{c}
  Patient Date of Birth \\
  \includegraphics[width=8cm,height=3.8cm]{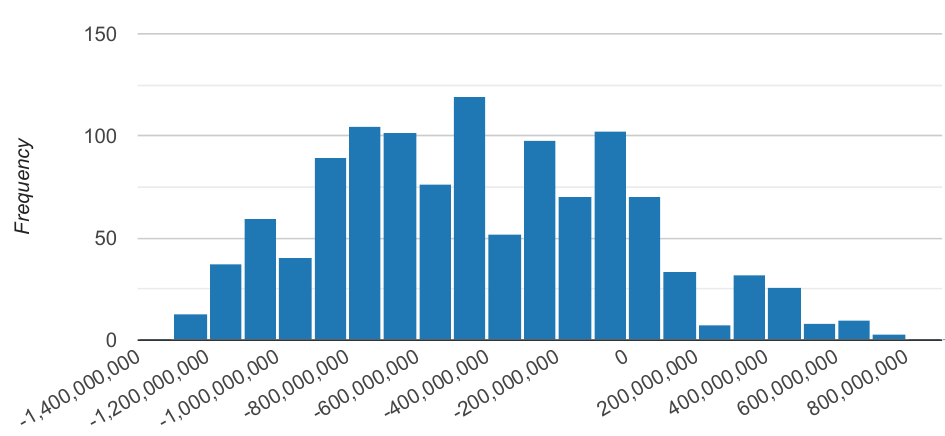} \\
  \vspace{0.01cm} \\
  Patient Health and Disability Decile \\
  \includegraphics[width=8cm,height=3.8cm]{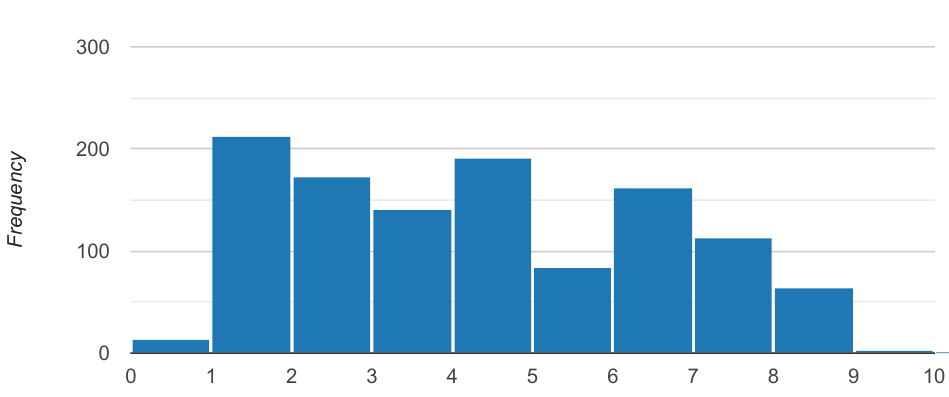} \\
  \end{tabular}
  \caption[]{Histogram analysis of the patient date of birth and patient health and disability decile metadata present in the multimodal dataset used in our experiments. Non-normal distribution is demonstrated in both types of data. Note that patient date of birth values are represented as timestamps.}
  \label{fig:histograms}
\end{figure}

We completed a histogram analysis of the DOB and HDD patient metadata to determine how the data should be represented in our deep learning experiments. This analysis showed that the DOB and HDD patient metadata categories did not exhibit normal distributions (see Figure \ref{fig:histograms}). Therefore, the patient DOB and HDD metadata used in our experiments were normalised using min max scaling as defined in Equation \ref{eq:norm}. 
\begin{equation}
  X^{'} = \frac{X - X_{min}}{X_{max}-X_{min}}
  \label{eq:norm}
\end{equation}

\noindent where $X$ is the data point and $X_{min}$ and $X_{max}$ are the minimum and maximum values present in the group respectively.

Patient gender was excluded from this analysis as there were only two possible values from the data provided, which we encoded as 0 (female) and 1 (male). 

\subsection{Gaussian Random Field Generation}

In this section we specify how GRF images were generated to encode patient metadata as spatially structured representations, enabling integration into deep learning-based multimodal chronic wound segmentation. This approach transforms numerical metadata into synthetic images that capture spatial correlations, ensuring compatibility with CNN architectures. The GRFs are constructed using a spectral synthesis approach, leveraging the properties of the Fourier transform to generate stochastic fields with controlled smoothness and variability \cite{leravalec2000fft}. 

Mathematically, a GRF is defined as a stochastic process $X(s)$ over a spatial domain $S$, such that any finite collection of values $\{X(s_1), X(s_2), ...,X(s_n)\}$ follows a multivariate normal distribution. Formally, for any finite set of points $\{s_1, s_2, ..., s_n\}$, the joint distribution of $X(s)$ is given by: 

\begin{equation}
  \left(X(s_1), X(s_2), ..., X(s_n)\right) \sim \mathcal{N}(\mu, K)
\end{equation}

\noindent where $\mu$ is the mean function $\mathbb{E}[X(s)]$ and $K(s,t)$ is the covariance function that defines the dependency between the points in the field. The covariance function must be positive semidefinite to ensure a valid Guassian process. 

The spatial structure of a GRF is characterised by its power spectrum $P_k$, which governs the correlation length and smoothness of the field. In this study, the power spectrum is dynamically determined based on the normalised patient metadata values. The power spectrum is computed as follows: 

\begin{equation}
  P_k = -\left| i + f \right|
\end{equation}

\noindent where $i$ is an integer component that defines the global structure and smoothness of the GRF, and $f$ is the fraction of the normalised metadata value. For binary nominal categorical variables such as gender, $f$ is set to zero, ensuring that the GRFs remain distinct for each category. Lower values of $P_k$ result in highly fragmented structures, whereas higher values produce smoother GRFs. 

The GRF generation process follows a spectral synthesis approach, which constructs a two-dimensional Gaussian noise field in the frequency domain using the Fast Fourier Transform (FFT). This noise field is modulated by the power spectrum $P_k$ to introduce spatial correlations, ensuring that local variations follow the prescribed smoothness constraints. The amplitude of the Fourier components is modified using the following equation: 

\begin{equation}
  A(k_x,k_y) = \sqrt{P_k(k_x, k_y)}
\end{equation}

\noindent where $k_x, k_y$ are the frequency components in the Fourier domain. This transformation ensures that low-frequency components are dominant within the field, generating large-scale spatial structures. High-frequency components contribute fine-grained details, but their influence diminishes as $P_k$ increases. 

Once the spectral domain representation is obtained, the inverse FFT is applied to transform the frequency-modulated field back into the spatial domain. The result is a two-dimensional GRF, expressed as a greyscale image, where the pixel intensities correspond to the metadata-driven stochastic process. 

\begin{figure}[!h]
  \centering
  \begin{tabular}{cc}
  (a) & (b) \\
  \includegraphics[width=3.7cm,height=3cm]{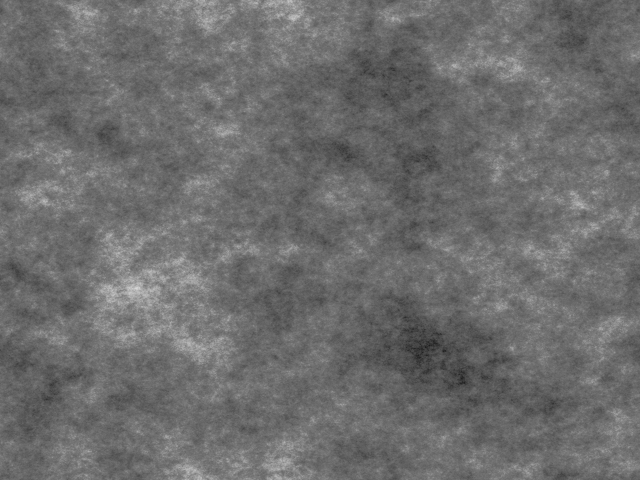} & 
  \includegraphics[width=3.7cm,height=3cm]{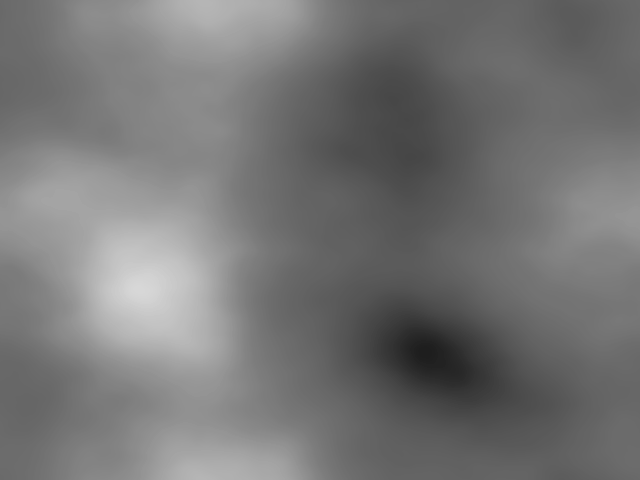} \\
  (c) & (d) \\
  \includegraphics[width=3.7cm,height=3cm]{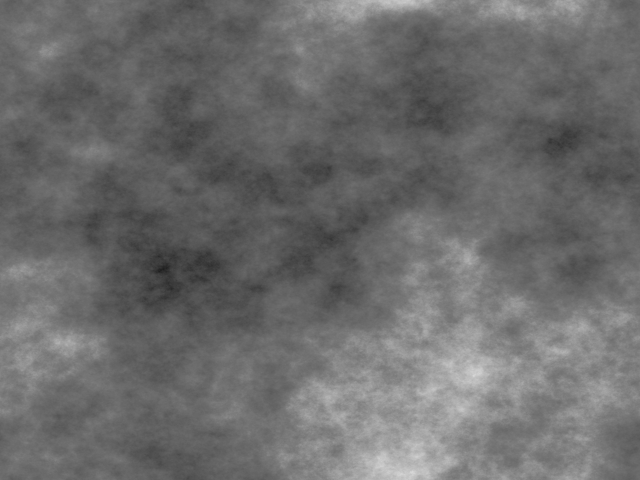} & 
  \includegraphics[width=3.7cm,height=3cm]{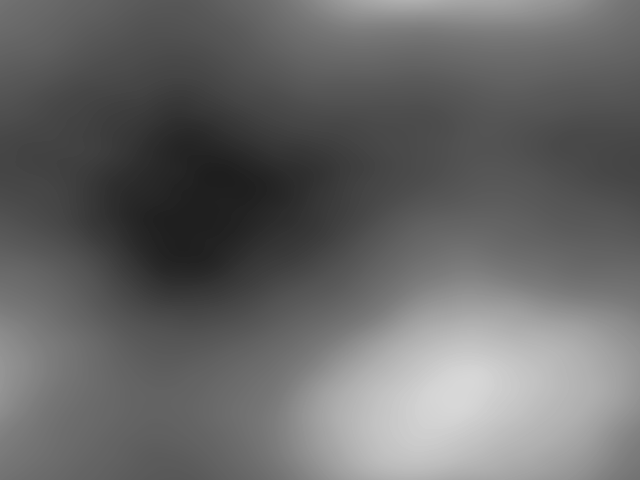} \\
  (e) & (f) \\
  \includegraphics[width=3.7cm,height=3cm]{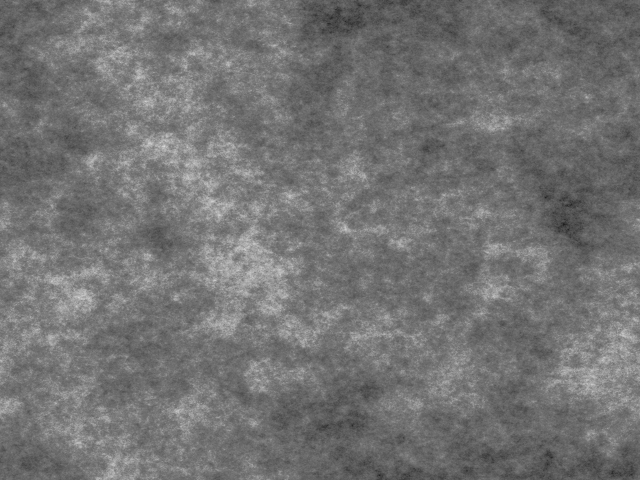} & 
  \includegraphics[width=3.7cm,height=3cm]{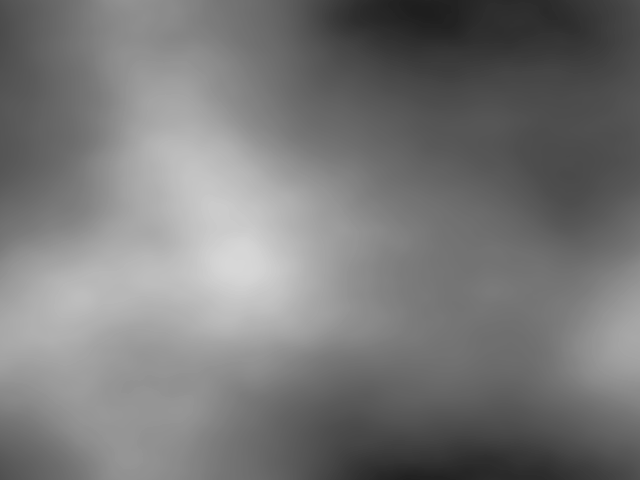} \\
  \end{tabular}
  \caption[]{Illustration of the types of Gaussian random fields generated for use in our multimodal chronic wound segmentation experiments. Examples (a), (c), and (e) were generated with an $i$ value of 2, and examples (b), (d), and (f) were generated using an $i$ value of 5. The top row examples were generated using DOB, the second row examples were generated using gender, and the third row examples were generated using HDD.}
  \label{fig:grf}
\end{figure}

All experiments used one of two types of GRF, where $i = 2$, or $i = 5$, as visualised in Figure \ref{fig:grf} (a, b), respectively. The difference lies in their structural complexity and smoothness, where GRFs generated with $i = 2$ exhibit greater fragmentation with localised variations and more fine granularity, making them suitable for capture of highly dynamic spatial dependencies. In contrast, GRFs with $i = 5$ result in smoother and more uniform structures, which are better suited to encode broader trends and continuous variations in the metadata. The choice of $i$ directly impacts the level of details within the GRF representation, with lower values producing noisier patterns and higher values resulting in more coherent spatial structures. 

To ensure reproducibility and consistency across experiments, a fixed random seed is assigned for each metadata category: \textbf{DOB} (76539635), \textbf{gender} (88118546), and \textbf{HDD} (41094303). Each GRF is stored as a $480\times640$ pixel greyscale image and merged with the RGB wound image tensors to form a four-channel input. This approach allows CNNs to extract metadata-related features without requiring direct numerical encoding, preserving an image-based representation paradigm within the segmentation pipeline. The generated GRFs act as an abstract representation of metadata, allowing the model to leverage spatial dependencies while maintaining compatibility with standard convolutional architectures. 



\subsection{Metrics}
To evaluate the performance of our multimodal chronic wound segmentation models we utilise a range of widely used metrics. Intersection over union (IoU) and Dice similarity coefficient (DSC) were used as the metrics for ascertaining segmentation model accuracy. DSC was selected for its representation as the harmonic mean of precision and recall, providing a more balanced evaluation of false positive and false negative prediction results. The mathematical expressions for IoU and DSC are shown in Equations \ref{eq:iou} and \ref{eq:dice} respectively. 

\begin{equation}
  IoU = \frac{|X \cap  Y|}{|X| + |Y|}
  \label{eq:iou}
\end{equation}

\begin{equation}
  DSC = 2 * \frac{|X \cap  Y|}{|X| + |Y| - |X \cap Y|}
  \label{eq:dice}
\end{equation}
\noindent
where $X$ and $Y$ indicate the ground truth and predicted masks, respectively. 

We also use two statistical hypothesis test metrics to provide an improved understanding of Type I and Type II errors found in deep learning segmentation algorithm results: False Positive Error (FPE) as detailed in Equation (\ref{eq:fpe}), and False Negative Error (FNE) as detailed in Equation (\ref{eq:fne}).

\begin{equation}
  FPE = \frac{FP}{FP + TN}
  \label{eq:fpe}
\end{equation}

\begin{equation}
  FNE = \frac{FN}{FN + TP}
  \label{eq:fne}
\end{equation}

\noindent
where $FP$ represents the total number of false positive predictions, $TN$ is the total number of true negative predictions, and $FN$ is the total number of false negative predictions. 






\subsection{Chronic Wound Datasets}
For our multimodal segmentation experiments we use a new private chronic wound dataset obtained from Lancashire Teaching Hospitals NHS Foundation Trust, UK. The use of this dataset was approved by the NHS Research Ethics Committee and the Health Research Authority (REF: SE-281). Written informed consent was obtained from all participating patients. This new multimodal dataset was collected between January 2023 and December 2023 during patient clinical appointments. A total of 1142 chronic wound images were captured using three digital cameras: a Kodak DX4530 (5 megapixel), a Nikon COOLPIX P100 (10.3 megapixel), and a Nikon D3300 (24.2 megapixel). Auto-focus was used during capture with no zoom or macro functions active, with an aperture setting of f/2.8 at a distance of approximately 30–40 cm from the wound surface. Natural lighting in the hospital settings was used instead of a flash. All chronic wound images were acquired by medical photographers whose specialisation is chronic wounds, all with more than 5 years professional clinical experience. Patient data was captured by clinicians who recorded the patient's date of birth (DOB), gender, and post code. Ground truth masks were generated using the HarDNet-CWS segmentation model proposed by \cite{cassidy2024cws}. Therefore, all training and validation experiments completed in this study are to be considered weakly supervised. The new chronic wound dataset is used in our experiments for training and validation. For testing, we use the DFUC 2022 test set which comprises 2000 DFU wound images and associated ground truth masks \cite{kendrick2025segmentation}. The DFUC 2022 test set does not have any associated metadata. 
A summary of the composition of characteristics of the training set is summarised in Table \ref{table:train_set_characteristics}. Note that multiple wound images may have been collected for a single patient during a hospital appointment. Such cases may include images of a single wound or multiple wounds, and may include images of the same wound captured at different angles and distances. 

    

\begin{table}[!h]
  \centering
  \renewcommand{\arraystretch}{1.0}
  \caption{Baseline characteristics for the multimodal training dataset comprising chronic wound images with corresponding patient metadata and weakly supervised ground truth masks. Note that in two cases it was not possible to exactly identify the wound type - ``arterial or venous" and ``venous or pressure". $^*$ - with fungal component.}
  \scalebox{0.94}{
    \label{table:train_set_characteristics}
    \begin{tabular}{|l|l|}
    \hline
    Category                                           & Number \\ \hline \hline
    No. of wound images                                & 1142   \\
    No. of DFU wound images                            & 1111   \\
    No. of venous wound images                         & 13     \\
    No. of arterial wound images                       & 12     \\
    No. of pressure wound images                       & 1      \\
    No. of dermatoliposclerosis wound images           & 1      \\
    No. of bacterial infection wound images$^*$        & 1      \\
    No. of ulcer on necrobiosis lipoidica wound images & 1      \\
    No. of arterial or venous wound images             & 1      \\
    No. of venous or pressure wound images             & 1      \\
    No. of patients                                    & 308    \\
    No. of appointments                                & 94     \\
    No. of male patients                               & 229    \\
    No. of female patients                             & 79     \\
    Median patient age                                 & 70     \\
    Median male patient age                            & 69     \\
    Median female patient age                          & 70     \\
    \hline
    \end{tabular}
  }
\end{table}





\subsection{Baseline}
We use the HarDNet-CWS network architecture as the basis of our multimodal experiments. HardDNet-CWS is the network architecture developed in our prior work in chronic wound segmentation \cite{cassidy2024cws}. HarDNet-CWS is a hybrid transformer segmentation architecture that uses traditional convolutional techniques in the encoder, and a vision transformer in the decoder. No augmentation or post-processing methods were used in any of our experiments. Pretrained weights were also not used. The baseline model was trained for 60 epochs with a batch size of 5 using the AdamW optimiser with a learning rate of 0.00001, an epsilon of 0.0000001, and a weight decay of 0.01. For the baseline experiment, only wound images were used for training, validation, and testing. 

\subsection{Hardware and Software Configuration}
The following hardware and software configuration was used in all our experiments: Debian GNU/Linux 10 (buster) operating system, AMD Ryzen 9 3900X 12-Core CPU, 128GB RAM, NVIDIA GeForce RTX 3090 24GB GPU. Models were trained with Pytorch 1.13.1 using Python 3.7.13.


\subsection{Training with Gaussian Random Fields using Tensor Merging}
\begin{figure}[!h]
  \centering
  \begin{tabular}{c}
  \includegraphics[width=6.8cm,height=4.3cm]{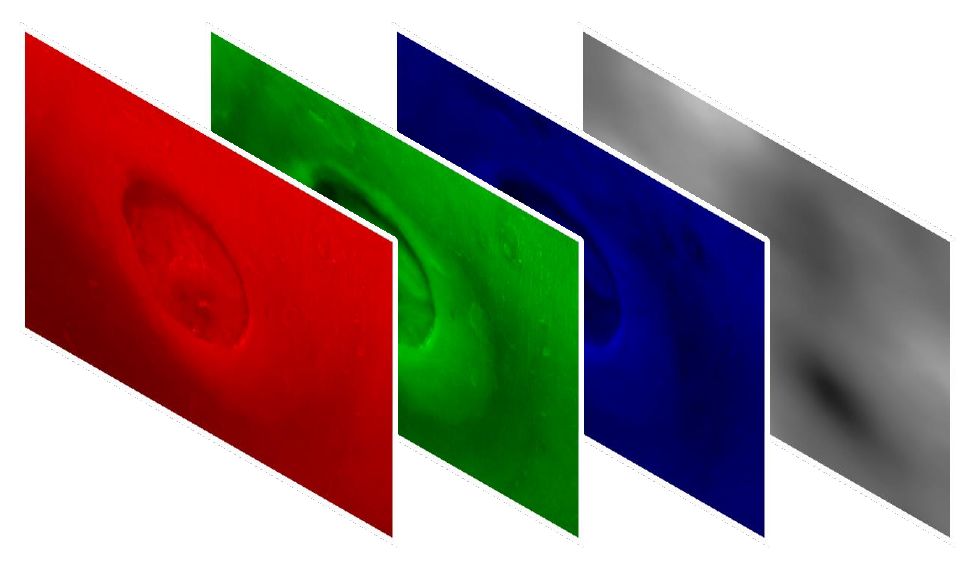} \\
  \end{tabular}
  \caption[]{Illustration of the tensor merging operation when merging wound image RGB tensors with single channel GRF tensors to produce new 4D tensors. The three RGB channels represent the wound image features, while the GRF channel is an abstract visual representation of a patient metadata item.}
  \label{fig:merge}
\end{figure}

Inspired by recent multi-colour space tensor merging experiments conducted by \cite{mcbride2024colour}, we experiment by introducing the GRF images into the training workflow by merging single channel GRFs with the RGB tensors representing the actual wound images. The tensor merging process was completed by merging the GRF single channel tensor onto the end of the RGB tensor which forms a new 4D tensor, as shown in Figure \ref{fig:merge}. Algorithm \ref{alg:merge} shows a pseudo code summary of the RGB and GRF tensor merging process. Early fusion was selected to ensure that inter-modal interactions occur throughout the network during training, allowing for richer feature representations to be learnt. 

\begin{algorithm}[!h]
  \caption{RGB+GRF tensor merging algorithm.}
  \label{alg:merge}
  \begin{algorithmic}[1]
    \Procedure{Tensor\_Merge}{$rgb\_image$, $grf\_image$}
    \State $rgb\_tensor \gets to\_tensor(rgb\_image)$
    \State $grf\_image \gets to\_greyscale(grf\_image)$
    \State $grf\_tensor \gets to\_tensor(grf\_image)$
    \State $4d\_tensor \gets merge([rgb\_tensor, grf\_tensor])$
    \State $4d\_tensor \gets normalise(4d\_tensor)$
    \State Return $4d\_tensor$
    \EndProcedure
  \end{algorithmic}
\end{algorithm}

\subsection{Average Merging of Prediction Masks}
\begin{figure}[!h]
  \centering
  \begin{tabular}{ccccc}
  (a) & (b) & (c) & (d) & (e) \\
  \includegraphics[width=1.38cm,height=1.1cm]{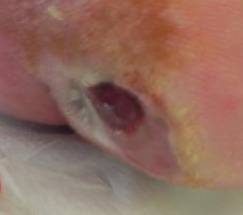} &
  \includegraphics[width=1.38cm,height=1.1cm]{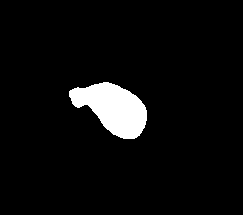} &
  \includegraphics[width=1.38cm,height=1.1cm]{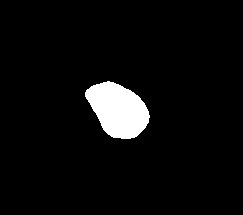} &
  \includegraphics[width=1.38cm,height=1.1cm]{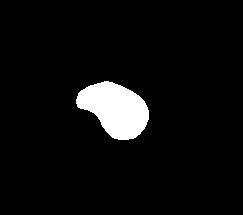} &
  \includegraphics[width=1.38cm,height=1.1cm]{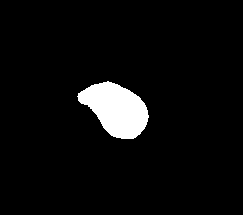} \\
  \end{tabular}
  \caption[]{Illustration of the prediction mask average merging process completed using distance transforms: (a) shows the original wound image, (b) is a mask generated by the model trained using DOB GRFs, (c) is a mask generated by the model trained using gender GRFs, (d) is a mask generated by the model trained using HDD GRFs, and (e) is the average merged mask. Note that images have been cropped for illustrative purposes.}
  \label{fig:dist_transforms}
\end{figure}

To further enhance prediction results, we complete a series of experiments whereby prediction masks are merged, using average merging, from predictions generated for the test set for the models trained using $i = 2$ and $i = 5$. Three sets of results are produced: (1) average merging of prediction results for the models trained using GRFs generated from DOB, gender, and HDD metadata where $i = 2$; (2) average merging of prediction results for the models trained using GRFs generated from DOB, gender, and HDD metadata where $i = 5$; and (3) average merging of prediction results for the models trained using GRFs generated from DOB, gender, and HDD metadata where $i = 2$ and $i = 5$. Prediction masks were averaged using the OpenCV distance transform method \cite{opencv2000}. The distance transform calculation is shown in Equation \ref{eq:dist_transform}. An example of prediction mask average merging using distance transforms is shown in Figure \ref{fig:dist_transforms}. 

\begin{equation}
  \begin{aligned}
    D_p(p) = \min_{q \in G} (d(p,q)+1(q)) \\
      1(q) =
      \begin{cases}
        0 & \text{if $q \in P$}\\
        \infty & \text{otherwise}\\
    \end{cases}
  \end{aligned}
  \label{eq:dist_transform}
\end{equation}

\noindent where $P$ is a set of points on grid $G$ ($P \subseteq G$), and associates to each grid location $q \in G$ the distance to the nearest point $p \in P$. 

\section{Results}
In this section we report on the results of inference for the baseline model, which was trained using only wound images, and the models trained using wound images with the patient metadata expressed as GRFs. 

\subsection{Baseline Results}
The training, validation, and test results for the baseline model are summarised in Table \ref{table:baseline}. 

\begin{table*}
    \centering
    \caption{Baseline results for the HarDNet-CWS model trained and validated using only wound images from the new multimodal dataset, and tested on the DFUC 2022 test set, which also comprises only wound images (image size = $640 \times 480$ pixels). IoU - intersection over union; DSC - Dice similarity coefficient; FPE - false positive error; FNE - false negative error. Note that no pretraining or post-processing was used in this experiment.}
    \label{table:baseline}
    \scalebox{0.98}{
    \begin{tabular}{|p{0.95cm}|p{1.4cm}|p{1.55cm}|p{1.5cm}|p{1.15cm}|p{1.25cm}|p{1.28cm}|p{1.25cm}|p{1.35cm}|p{1.0cm}|p{1.0cm}|}
        \hline
        Epoch & Train IoU & Train Loss & Train DSC & Val IoU & Val Loss & Val DSC & Test IoU & Test DSC & FPE     & FNE    \\ \hline \hline
        44    & 0.9346    & 0.0972     & 0.9658    & 0.5887  & 0.5246   & 0.7018  & 0.4670   & 0.5908   & 0.0169  & 0.3637 \\
        \hline
	\end{tabular}
	}
\end{table*}

\subsection{Gaussian Random Field Results}
The results for the chronic wound segmentation experiments, where the patient metadata expressed as single channel greyscale GRF images was included into the training workflow, are summarised in Table \ref{table:grf}. 

\begin{table*}
    \centering
    \caption{Results for the HarDNet-CWS model trained and validated using the new multimodal dataset, and tested on the DFUC 2022 test set (image size = $640 \times 480$ pixels) using GRFs. $i$ - default value of GRF power spectrum integer component; IoU - intersection over union; DSC - Dice similarity coefficient; FPE - false positive error; FNE - false negative error; DOB - date of birth; HDD - health and disability decile. Note that no pretraining, augmentation, or post-processing was used in these experiments.}
    \label{table:grf}
    \scalebox{0.86}{
    \begin{tabular}{|p{1.4cm}|p{0.23cm}|p{0.95cm}|p{1.4cm}|p{1.55cm}|p{1.55cm}|p{1.15cm}|p{1.25cm}|p{1.25cm}|p{1.25cm}|p{1.4cm}|p{0.98cm}|p{0.98cm}|}
        \hline
        Metadata & $i$ & Epoch & Train IoU & Train Loss & Train DSC & Val IoU & Val Loss & Val DSC & Test IoU & Test DSC & FPE    & FNE    \\ \hline
        \hline
        DOB      & 2   & 44    & 0.9323    & 0.1012     & 0.9647    & 0.5615  & 0.5539   & 0.6760  & 0.4699   & 0.5944   & 0.0169 & 0.3562 \\
        \hline
        DOB      & 5   & 54    & 0.9566    & 0.0670     & 0.9778    & 0.5654  & 0.5866   & 0.6842  & 0.4652   & 0.5894   & 0.0169 & 0.3475 \\
        \hline
        Gender   & 2   & 34    & 0.8810    & 0.1659     & 0.9359    & 0.5554  & 0.5277   & 0.6708  & 0.4703   & 0.5942   & 0.0169 & 0.3575 \\
        \hline
        Gender   & 5   & 36    & 0.8843    & 0.1635     & 0.9376    & 0.5832  & 0.5038   & 0.6976  & 0.4641   & 0.5889   & 0.0169 & 0.3621 \\
        \hline
        HDD      & 2   & 47    & 0.9454    & 0.0834     & 0.9718    & 0.5825  & 0.5657   & 0.6962  & \textbf{0.4706}   & \textbf{0.5946}   & \textbf{0.0169} & \textbf{0.3424} \\
        \hline
        HDD      & 5   & 48    & 0.9394    & 0.0893     & 0.9685    & 0.5672  & 0.5652   & 0.6819  & 0.4626   & 0.5873   & 0.0169 & 0.3465 \\
        \hline
	\end{tabular}
	}
\end{table*}

\subsection{Gaussian Random Field Experiment Predictions with Average Merging}
The results of the GRF experiments where predictions from different models, each trained and validated on a single GRF type and combined using average merging, are summarised in Table \ref{table:avg}. 

\begin{table}[!h]
    \centering
    \caption{Test results for the HarDNet-CWS models with averaged prediction masks. $i$ - default value of GRF power spectrum integer component; IoU - intersection over union; DSC - Dice similarity coefficient; FPE - false positive error; FNE - false negative error; DOB - date of birth; Gen - gender, HDD - health and disability decile. Note that no pretraining, augmentation, or post-processing was used in these experiments.}
    \label{table:avg}
    \scalebox{0.84}{
    \begin{tabular}{|p{2.65cm}|p{1.0cm}|p{1.0cm}|p{1.0cm}|p{1.0cm}|p{1.0cm}|}
    \hline
    Metadata    & $i$    & IoU             & DSC             & FPE             & FNE    \\ \hline \hline
    DOB+Gen+HDD & 2      & \textbf{0.4899} & \textbf{0.6128} & \textbf{0.0199} & 0.3209 \\ \hline
    DOB+Gen+HDD & 5      & 0.4841          & 0.6079          & 0.0199          & 0.3192 \\ \hline
    DOB+Gen+HDD & 2 \& 5 & 0.4897          & 0.6122          & 0.0254          & \textbf{0.3171} \\ \hline
    \end{tabular}
    }
\end{table}




\section{Discussion}
The results of models trained using RGB wound images and patient metadata expressed as GRFs (see Table \ref{table:grf}) indicate that for all models trained where $i = 2$, improvements in terms DSC were demonstrated when compared to the baseline results. The best overall performing GRF model was the HDD model ($i = 2$) which demonstrated improvements of 0.0036, 0.0038, and 0.0213 for IoU, DSC, and FNE respectively when compared to the baseline results. We also observe that in terms of FPE, the baseline and subsequent experiment results are unchanged, with a reported value of 0.0169. The IoU and DSC improvements for these experiments are marginal ($< 1\%$), whereas we would consider the improvement of FNE to be significant ($> 2\%$). 

For the experiments using averaged prediction masks from all model types (DOB, gender, and HDD - see Table \ref{table:avg}), we observe that the models trained with $i = 2$ provided the highest performance improvements. The averaged model predictions trained with $i = 2$ demonstrated improvements in terms of IoU (+0.0229) and DSC (+0.0220) when compared to the baseline results. When compared to the results for models trained on individual metadata categories (DOB, gender, and HDD - see Table \ref{table:grf}), improvements are observed in terms of IoU (+0.0193) and DSC (+0.0182) for the $i = 2$ models, and FNE (-0.0253) for the combined $i = 2$ and $i = 5$ models. When compared to the models trained on individual metadata categories (see Table \ref{table:grf}), these results suggest that increasing the number of patient metadata categories improves overall network performance in terms of IoU, DSC, and FNE. We suggest that these results are promising, considering that our experiments were conducted using metadata only for the training and validation sets. 

\begin{figure*}[!h]
  \centering
  \begin{tabular}{cccc}
  (a) & (b) & (c) & (d) \\
  \includegraphics[width=4.2cm,height=3.2cm]{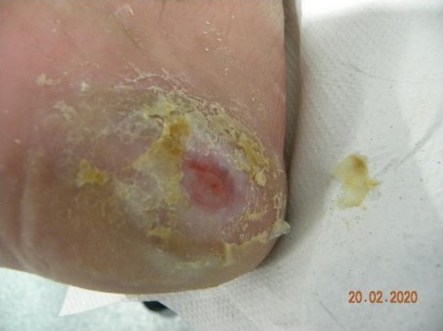} &
  \includegraphics[width=4.2cm,height=3.2cm]{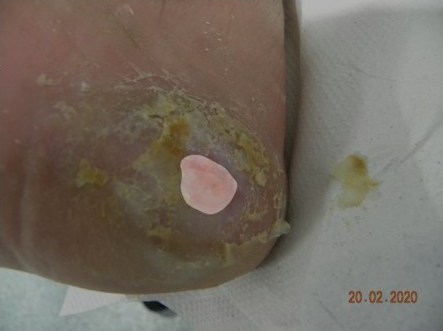} &
  \includegraphics[width=4.2cm,height=3.2cm]{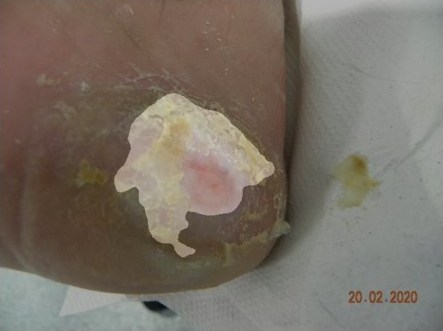} & 
  \includegraphics[width=4.2cm,height=3.2cm]{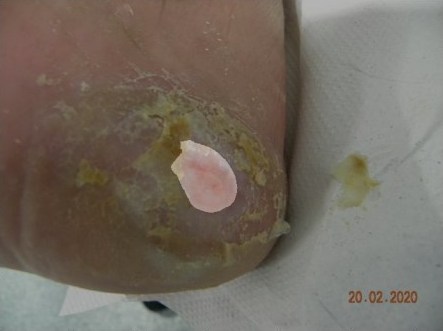} \\

  \includegraphics[width=4.2cm,height=3.2cm]{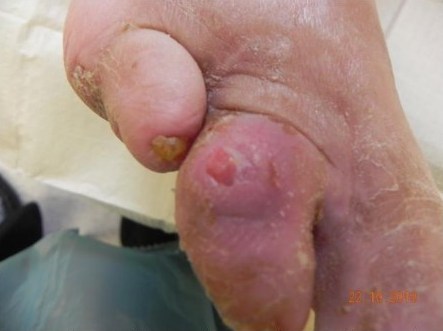} &
  \includegraphics[width=4.2cm,height=3.2cm]{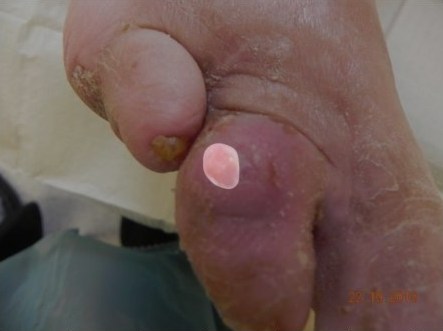} &
  \includegraphics[width=4.2cm,height=3.2cm]{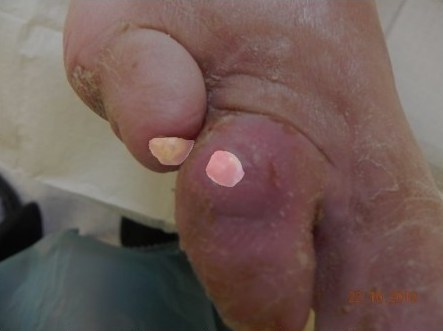} & 
  \includegraphics[width=4.2cm,height=3.2cm]{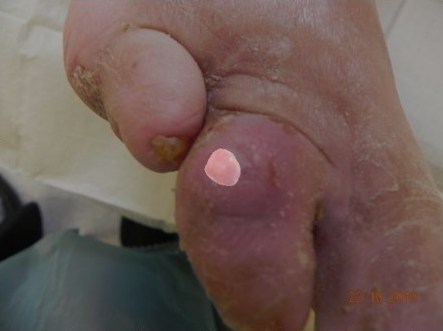} \\

  \includegraphics[width=4.2cm,height=3.2cm]{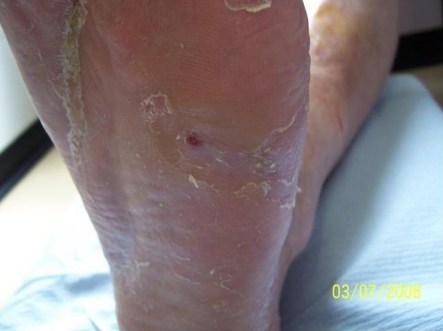} &
  \includegraphics[width=4.2cm,height=3.2cm]{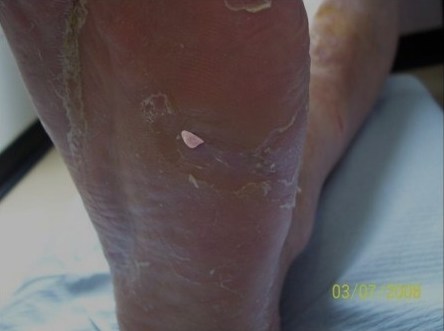} &
  \includegraphics[width=4.2cm,height=3.2cm]{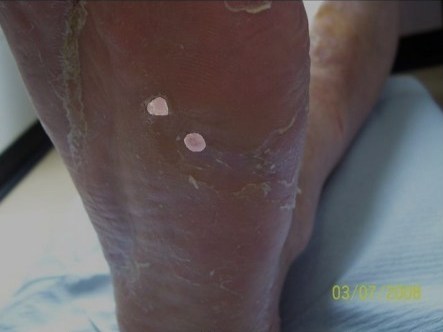} & 
  \includegraphics[width=4.2cm,height=3.2cm]{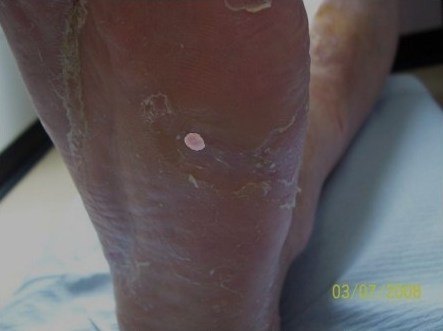} \\

  \includegraphics[width=4.2cm,height=3.2cm]{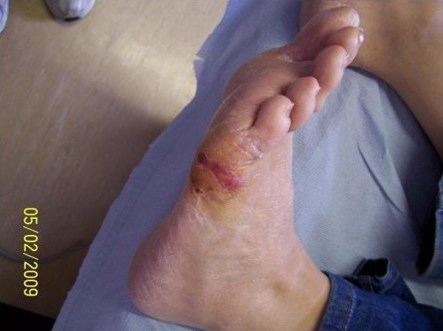} &
  \includegraphics[width=4.2cm,height=3.2cm]{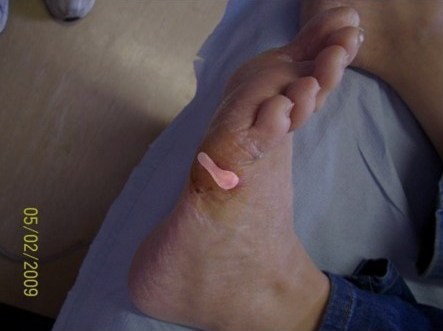} &
  \includegraphics[width=4.2cm,height=3.2cm]{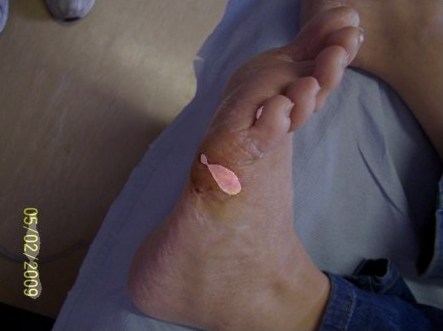} & 
  \includegraphics[width=4.2cm,height=3.2cm]{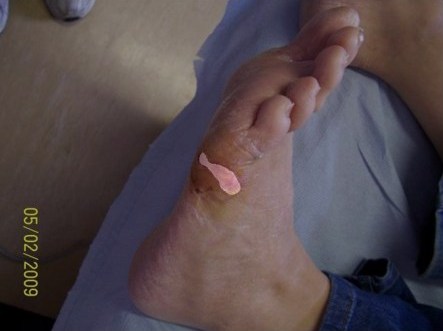} \\
  \end{tabular}
  \caption[]{Illustration showing four examples of where the averaged merged predictions from the $i = 2$ models show improved performance over the baseline results. Column (a) shows the original wound image, column (b) shows the ground truth mask, column (c) shows the baseline HarDNet-CWS prediction, and column (d) shows the averaged merged prediction for the $i = 2$ models.}
  \label{fig:improved}
\end{figure*}

A selection of baseline and averaged merged predictions from the $i = 2$ models is shown in Figure \ref{fig:improved}. 
The first row shows a case where the baseline result has incorrectly predicted parts of the dried skin surrounding the wound, whereas the averaged merged prediction is significantly closer to the ground truth mask. 
The second row shows a case where the wound region has been correctly predicted by both baseline and averaged merged predictions, however, the baseline has also incorrectly predicted a toenail as a wound region. 
The third row shows a case where the baseline and averaged merged predictions have correctly predicted the wound located on the plantar aspect, however, the baseline model has also incorrectly predicted on a region of dried skin. 
The fourth row shows a case where the baseline has predicted less of the wound region compared to the averaged merged predictions. 

The rationale for using GRF images in our multimodal chronic wound segmentation experiments was to ensure that all input data, and therefore learnable features, remained strictly within the confines of the computer vision domain. That is, all learnable features could be derived from imaging domains, regardless of the nature of the data (image or numerical data). Generation of synthetic images to represent non-image data was previously proposed by \cite{sharma2019deepinsight}, named DeepInsight, for classification tasks. They converted feature vectors derived from text values to a feature matrix which is used to generate an image. However, their approach generated sparse pixel patterns in images that CNNs may find difficult to learn features from and may not scale well to higher resolutions. Their method was also not used in terms of multimodality. Our approach is capable of generating dense features at any resolution regardless of the complexity of the input value, and has been demonstrated in a multimodal segmentation architecture. Additionally, GRF images exhibit complex texture features, of which CNNs naturally exhibit a learning bias towards \cite{chung2023texture}, which provides a further potential advantage over the method proposed by \cite{sharma2019deepinsight}. 


Collection and curation of patient medical datasets presents numerous significant challenges. Obtaining patient metadata can be time consuming and involves stringent ethical procedures. Making such datasets public also poses risks regarding patient anonymity. Distribution of patient metadata as abstract representations, as proposed in the present paper, may be a way to circumvent many of the challenges associated with the distribution of patient data and may encourage other research groups to share much needed data with the research community. 

We speculate that the improvement of IoU and DSC test metrics for the ensemble GRF experiments indicate three possible scenarios. First, the GRF features may be acting as a form of augmentation during training. The second possibility is that the model may be learning GRF features in a manner that associates wound features with GRF features. The third possibility is that the GRF features are acting as both a form of augmentation and as wound-associating features. 

Prior works in chronic wound deep learning segmentation research have shown a disparity between lab-based metrics and measures derived from expert qualitative assessment. Cassidy et al. \cite{cassidy2024cws} found that lab-based metrics were less favourable when compared to expert clinical judgement in chronic wound segmentation tasks, indicating that the lab-based metrics were not fully reflective of clinical assessment. It is therefore important for future works involving the use of patient data in chronic wound deep learning studies to evaluate differences between such quantitative and qualitative measures to ensure real-world validity of proposed models. This important aspect of chronic wound segmentation will be explored further in our future multimodal experiments. Access to clinical experts can be challenging, however, the absence of such expert assessment should be viewed as a limitation of any deep learning research involving medical imaging, including the present study. 

Another limitation of the present study involves the limited size of the training set. The data used in our study represents both chronic wound images and associated patient metadata. With just 1142 chronic wound images, collected from 308 patients, this data is limited in size and may also exhibit bias due to the limited number of patients. However, despite these limitations, factors such as images obtained at different appointments and at different angles and orientations may to some extent circumvent the apparent biases. We also observe other imbalances in the training set, particularly in terms of male and female distribution, and to a lesser extent, median male and female patient age. Our research group is currently in the process of collecting additional multimodal data from hospitals both in the UK and internationally with a future focus on utilising such datasets in upcoming research which will build on the methods presented in the current study. 


Preprocessing methods have been shown to be valuable techniques when training deep learning models in medical imaging domains \cite{dipto2023similarity, jaworek2023hair, pewton2024dca, okafor2024preproc}. However, we do not report on such methods in the present paper, as those methods did not provide performance benefits over the current state-of-the-art for the test set used in our experiments. 


The multimodal segmentation experiments conducted in the present paper were weakly supervised. Therefore, we present these findings as preliminary research which may be used as a basis of comparison in future studies when more patient data becomes available, allowing for more extensive multimodal segmentation studies. We encourage other research groups working in multimodal medical imaging deep learning tasks to further explore the concepts presented in the present paper.

\section{Conclusion}
In this work we propose the use of GRFs generated using patient metadata for use in a chronic wound segmentation workflow. Our results demonstrate that via an ensemble of models, each trained on different GRFs generated from different patient metadata categories, we were able to outperform the baseline experiment results (without GRFs) in terms of IoU (+0.0229) and DSC (+0.0220). These experimental results were achieved using weak supervision for the training data, in addition to the use of a training set that was significantly smaller when compared to the test set (train set = 1142; test set = 2000). Our approach allows for the introduction of patient data into multimodal CNN models with minimal adjustments to the architecture design. Additionally, we demonstrate that the test results can be improved with the use of GRFs where only the training data has associated metadata. Our findings indicate that the use of GRFs as an abstract representation of patient metadata is a viable option in deep learning training workflows for segmentation, with potential utility in classification and localisation tasks.

\section*{Acknowledgments}
We would like to thank clinicians at Lancashire Teaching Hospitals NHS Foundation Trust, UK who collected the multimodal chronic wound dataset used in our chronic wound segmentation experiments.

\bibliographystyle{model2-names.bst}
\bibliography{Ref}

\begin{thebibliography}{61}
\expandafter\ifx\csname natexlab\endcsname\relax\def\natexlab#1{#1}\fi
\providecommand{\url}[1]{\texttt{#1}}
\providecommand{\href}[2]{#2}
\providecommand{\path}[1]{#1}
\providecommand{\DOIprefix}{doi:}
\providecommand{\ArXivprefix}{arXiv:}
\providecommand{\URLprefix}{URL: }
\providecommand{\Pubmedprefix}{pmid:}
\providecommand{\doi}[1]{\href{http://dx.doi.org/#1}{\path{#1}}}
\providecommand{\Pubmed}[1]{\href{pmid:#1}{\path{#1}}}
\providecommand{\bibinfo}[2]{#2}
\ifx\xfnm\relax \def\xfnm[#1]{\unskip,\space#1}\fi
\bibitem[{Anisuzzaman et~al.(2022)Anisuzzaman, Patel, Rostami, Niezgoda,
  Gopalakrishnan and Yu}]{anisuzzaman2022multimodal}
\bibinfo{author}{Anisuzzaman, D.M.}, \bibinfo{author}{Patel, Y.},
  \bibinfo{author}{Rostami, B.}, \bibinfo{author}{Niezgoda, J.},
  \bibinfo{author}{Gopalakrishnan, S.}, \bibinfo{author}{Yu, Z.},
  \bibinfo{year}{2022}.
\newblock \bibinfo{title}{Multi-modal wound classification using wound image
  and location by deep neural network}.
\newblock \bibinfo{journal}{Scientific Reports} \bibinfo{volume}{12}.
\newblock \DOIprefix\doi{10.1038/s41598-022-21813-0}.
\bibitem[{Apelqvist et~al.(1993)Apelqvist, Larsson and
  Agardh}]{apelqvist1993prognosis}
\bibinfo{author}{Apelqvist, J.}, \bibinfo{author}{Larsson, J.},
  \bibinfo{author}{Agardh, C.D.}, \bibinfo{year}{1993}.
\newblock \bibinfo{title}{Long-term prognosis for diabetic patients with foot
  ulcers}.
\newblock \bibinfo{journal}{Journal of Internal Medicine}
  \bibinfo{volume}{233}, \bibinfo{pages}{485--491}.
\newblock \DOIprefix\doi{https://doi.org/10.1111/j.1365-2796.1993.tb01003.x}.
\bibitem[{Bader(2008)}]{bader2008diabetic}
\bibinfo{author}{Bader, M.S.}, \bibinfo{year}{2008}.
\newblock \bibinfo{title}{Diabetic foot infection}.
\newblock \bibinfo{journal}{American family physician} \bibinfo{volume}{78}.
\bibitem[{Boulton et~al.(2005)Boulton, Vileikyte, Ragnarson-Tennvall and
  Apelqvist}]{boulton2005burden}
\bibinfo{author}{Boulton, A.J.}, \bibinfo{author}{Vileikyte, L.},
  \bibinfo{author}{Ragnarson-Tennvall, G.}, \bibinfo{author}{Apelqvist, J.},
  \bibinfo{year}{2005}.
\newblock \bibinfo{title}{The global burden of diabetic foot disease}.
\newblock \bibinfo{journal}{The Lancet} \bibinfo{volume}{366},
  \bibinfo{pages}{1719--1724}.
\newblock \URLprefix
  \url{https://www.sciencedirect.com/science/article/pii/S0140673605676982},
  \DOIprefix\doi{https://doi.org/10.1016/S0140-6736(05)67698-2}.
\bibitem[{Bradski(2000)}]{opencv2000}
\bibinfo{author}{Bradski, G.}, \bibinfo{year}{2000}.
\newblock \bibinfo{title}{{The OpenCV Library}}.
\newblock \bibinfo{journal}{Dr. Dobb's Journal of Software Tools}
  \bibinfo{volume}{25}, \bibinfo{pages}{120,122--125}.
\bibitem[{Brownrigg et~al.(2013)Brownrigg, Apelqvist, Bakker, Schaper and
  Hinchliffe}]{brownrigg2013pad}
\bibinfo{author}{Brownrigg, J.}, \bibinfo{author}{Apelqvist, J.},
  \bibinfo{author}{Bakker, K.}, \bibinfo{author}{Schaper, N.},
  \bibinfo{author}{Hinchliffe, R.}, \bibinfo{year}{2013}.
\newblock \bibinfo{title}{Evidence-based management of pad \& the diabetic
  foot}.
\newblock \bibinfo{journal}{European Journal of Vascular and Endovascular
  Surgery} \bibinfo{volume}{45}, \bibinfo{pages}{673--681}.
\newblock \URLprefix
  \url{https://www.sciencedirect.com/science/article/pii/S1078588413001366},
  \DOIprefix\doi{https://doi.org/10.1016/j.ejvs.2013.02.014}.
\bibitem[{Brüngel et~al.(2025)Brüngel, Kendrick, Cassidy, Bracke, Friedrich,
  Reeves, Pappachan and Yap}]{brungel2025overview}
\bibinfo{author}{Brüngel, R.}, \bibinfo{author}{Kendrick, C.},
  \bibinfo{author}{Cassidy, B.}, \bibinfo{author}{Bracke, B.},
  \bibinfo{author}{Friedrich, C.}, \bibinfo{author}{Reeves, N.},
  \bibinfo{author}{Pappachan, J.}, \bibinfo{author}{Yap, M.H.},
  \bibinfo{year}{2025}.
\newblock \bibinfo{title}{Diabetic Foot Ulcer Grand Challenge 2024: Overview
  and Baseline Methods}.
\newblock pp. \bibinfo{pages}{109--124}.
\newblock \DOIprefix\doi{10.1007/978-3-031-80871-5_10}.
\bibitem[{Cassidy et~al.(2023)Cassidy, Kendrick, Reeves, Pappachan, O'Shea,
  Chandrabalan and Yap}]{cassidy2023summary}
\bibinfo{author}{Cassidy, B.}, \bibinfo{author}{Kendrick, C.},
  \bibinfo{author}{Reeves, N.}, \bibinfo{author}{Pappachan, J.},
  \bibinfo{author}{O'Shea, C.}, \bibinfo{author}{Chandrabalan, V.V.},
  \bibinfo{author}{Yap, M.H.}, \bibinfo{year}{2023}.
\newblock \bibinfo{title}{Diabetic foot ulcer grand challenge 2022 summary}.
\newblock \DOIprefix\doi{10.1007/978-3-031-26354-5_10}.
\bibitem[{Cassidy et~al.(2022)Cassidy, Kendrick, Reeves, Pappachan, O’Shea,
  Armstrong and Yap}]{cassidy2021eval}
\bibinfo{author}{Cassidy, B.}, \bibinfo{author}{Kendrick, C.},
  \bibinfo{author}{Reeves, N.}, \bibinfo{author}{Pappachan, J.},
  \bibinfo{author}{O’Shea, C.}, \bibinfo{author}{Armstrong, D.},
  \bibinfo{author}{Yap, M.H.}, \bibinfo{year}{2022}.
\newblock \bibinfo{title}{Diabetic Foot Ulcer Grand Challenge 2021: Evaluation
  and Summary}.
\newblock pp. \bibinfo{pages}{90--105}.
\newblock \DOIprefix\doi{10.1007/978-3-030-94907-5_7}.
\bibitem[{Cassidy et~al.(2024)Cassidy, Mcbride, Kendrick, Reeves, Pappachan,
  Fernandez, Chacko, Brüngel, Friedrich, Alotaibi, AlWabel, Alderwish, Lai and
  Yap}]{cassidy2024cws}
\bibinfo{author}{Cassidy, B.}, \bibinfo{author}{Mcbride, C.},
  \bibinfo{author}{Kendrick, C.}, \bibinfo{author}{Reeves, N.D.},
  \bibinfo{author}{Pappachan, J.M.}, \bibinfo{author}{Fernandez, C.J.},
  \bibinfo{author}{Chacko, E.}, \bibinfo{author}{Brüngel, R.},
  \bibinfo{author}{Friedrich, C.M.}, \bibinfo{author}{Alotaibi, M.},
  \bibinfo{author}{AlWabel, A.A.}, \bibinfo{author}{Alderwish, M.},
  \bibinfo{author}{Lai, K.Y.}, \bibinfo{author}{Yap, M.H.},
  \bibinfo{year}{2024}.
\newblock \bibinfo{title}{An enhanced harmonic densely connected hybrid
  transformer network architecture for chronic wound segmentation utilising
  multi-colour space tensor merging}.
\newblock \bibinfo{journal}{arXiv preprint arXiv:2410.03359} .
\bibitem[{Chammas et~al.(2016)Chammas, Hill and Edmonds}]{chammas2016mortality}
\bibinfo{author}{Chammas, N.K.}, \bibinfo{author}{Hill, R.L.R.},
  \bibinfo{author}{Edmonds, M.E.}, \bibinfo{year}{2016}.
\newblock \bibinfo{title}{Increased mortality in diabetic foot ulcer patients:
  The significance of ulcer type}.
\newblock \bibinfo{journal}{Journal of Diabetes Research}
  \bibinfo{volume}{2016}, \bibinfo{pages}{2879809}.
\newblock \URLprefix
  \url{https://onlinelibrary.wiley.com/doi/abs/10.1155/2016/2879809},
  \DOIprefix\doi{https://doi.org/10.1155/2016/2879809}.
\bibitem[{Chin et~al.(2024)Chin, Lee, Sia and Hong}]{chin2024mortality}
\bibinfo{author}{Chin, B.}, \bibinfo{author}{Lee, P.}, \bibinfo{author}{Sia,
  C.H.}, \bibinfo{author}{Hong, C.C.}, \bibinfo{year}{2024}.
\newblock \bibinfo{title}{Diabetic foot ulcer is associated with
  cardiovascular-related mortality and morbidity – a systematic review and
  meta-analysis of 8062 patients}.
\newblock \bibinfo{journal}{Endocrine} \bibinfo{volume}{84},
  \bibinfo{pages}{1--12}.
\newblock \DOIprefix\doi{10.1007/s12020-024-03696-5}.
\bibitem[{Chung and Park(2023)}]{chung2023texture}
\bibinfo{author}{Chung, H.}, \bibinfo{author}{Park, K.}, \bibinfo{year}{2023}.
\newblock \bibinfo{title}{Shape Prior is Not All You Need: Discovering Balance
  Between Texture and Shape Bias in CNN}.
\newblock pp. \bibinfo{pages}{491--506}.
\newblock \DOIprefix\doi{10.1007/978-3-031-26284-5_30}.
\bibitem[{Codella et~al.(2019)Codella, Rotemberg, Tschandl, Celebi, Dusza,
  Gutman, Helba, Kalloo, Liopyris, Marchetti, Kittler and
  Halpern}]{codella2019isic}
\bibinfo{author}{Codella, N.}, \bibinfo{author}{Rotemberg, V.},
  \bibinfo{author}{Tschandl, P.}, \bibinfo{author}{Celebi, M.E.},
  \bibinfo{author}{Dusza, S.}, \bibinfo{author}{Gutman, D.},
  \bibinfo{author}{Helba, B.}, \bibinfo{author}{Kalloo, A.},
  \bibinfo{author}{Liopyris, K.}, \bibinfo{author}{Marchetti, M.},
  \bibinfo{author}{Kittler, H.}, \bibinfo{author}{Halpern, A.},
  \bibinfo{year}{2019}.
\newblock \bibinfo{title}{Skin lesion analysis toward melanoma detection 2018:
  A challenge hosted by the international skin imaging collaboration (isic)}.
\newblock \DOIprefix\doi{10.48550/arXiv.1902.03368}.
\bibitem[{Costa et~al.(2017)Costa, Cardoso, Procópio, Navarro, Dardik and
  Cisneros}]{costa2017mortality}
\bibinfo{author}{Costa, R.}, \bibinfo{author}{Cardoso, N.},
  \bibinfo{author}{Procópio, R.}, \bibinfo{author}{Navarro, T.},
  \bibinfo{author}{Dardik, A.}, \bibinfo{author}{Cisneros, L.},
  \bibinfo{year}{2017}.
\newblock \bibinfo{title}{Diabetic foot ulcer carries high amputation and
  mortality rates, particularly in the presence of advanced age, peripheral
  artery disease and anemia}.
\newblock \bibinfo{journal}{Diabetes \& Metabolic Syndrome: Clinical Research
  \& Reviews} \bibinfo{volume}{11}.
\newblock \DOIprefix\doi{10.1016/j.dsx.2017.04.008}.
\bibitem[{Dasari et~al.(2021)Dasari, Jiang, Skochdopole, Chung, Reece,
  Vorstenbosch and Winocour}]{dasari2021wound}
\bibinfo{author}{Dasari, N.}, \bibinfo{author}{Jiang, A.},
  \bibinfo{author}{Skochdopole, A.}, \bibinfo{author}{Chung, J.},
  \bibinfo{author}{Reece, E.}, \bibinfo{author}{Vorstenbosch, J.},
  \bibinfo{author}{Winocour, S.}, \bibinfo{year}{2021}.
\newblock \bibinfo{title}{Updates in diabetic wound healing, inflammation, and
  scarring}.
\newblock \bibinfo{journal}{Seminars in Plastic Surgery} \bibinfo{volume}{35}.
\newblock \DOIprefix\doi{10.1055/s-0041-1731460}.
\bibitem[{Dipto et~al.(2023)Dipto, Cassidy, Kendrick, Reeves, Pappachan,
  Chandrabalan and Yap}]{dipto2023similarity}
\bibinfo{author}{Dipto, I.}, \bibinfo{author}{Cassidy, B.},
  \bibinfo{author}{Kendrick, C.}, \bibinfo{author}{Reeves, N.},
  \bibinfo{author}{Pappachan, J.}, \bibinfo{author}{Chandrabalan, V.},
  \bibinfo{author}{Yap, M.H.}, \bibinfo{year}{2023}.
\newblock \bibinfo{title}{Quantifying the Effect of Image Similarity on
  Diabetic Foot Ulcer Classification}.
\newblock pp. \bibinfo{pages}{1--18}.
\newblock \DOIprefix\doi{10.1007/978-3-031-26354-5_1}.
\bibitem[{Fan et~al.(2024)Fan, Huang, Liu, Xie, Yang, Liang and
  Ding}]{fan2024risk}
\bibinfo{author}{Fan, Z.}, \bibinfo{author}{Huang, J.}, \bibinfo{author}{Liu,
  Y.}, \bibinfo{author}{Xie, H.}, \bibinfo{author}{Yang, Q.},
  \bibinfo{author}{Liang, Y.}, \bibinfo{author}{Ding, H.},
  \bibinfo{year}{2024}.
\newblock \bibinfo{title}{Multifactorial analysis of risk factors for foot
  ulcers in patients with neurovascular complications of diabetes}.
\newblock \bibinfo{journal}{Frontiers in Endocrinology} \bibinfo{volume}{15}.
\newblock \DOIprefix\doi{10.3389/fendo.2024.1399924}.
\bibitem[{Franks et~al.(2016)Franks, Barker, Collier, Gethin, Haesler, Jawien,
  Läuchli, Mosti, Probst and Weller}]{franks2016venous}
\bibinfo{author}{Franks, P.}, \bibinfo{author}{Barker, J.},
  \bibinfo{author}{Collier, M.}, \bibinfo{author}{Gethin, G.},
  \bibinfo{author}{Haesler, E.}, \bibinfo{author}{Jawien, A.},
  \bibinfo{author}{Läuchli, S.}, \bibinfo{author}{Mosti, G.},
  \bibinfo{author}{Probst, S.}, \bibinfo{author}{Weller, C.},
  \bibinfo{year}{2016}.
\newblock \bibinfo{title}{Management of patients with venous leg ulcers:
  Challenges and current best practice}.
\newblock \bibinfo{journal}{Journal of Wound Care} \bibinfo{volume}{25},
  \bibinfo{pages}{S1--S67}.
\newblock \DOIprefix\doi{10.12968/jowc.2016.25.Sup6.S1}.
\bibitem[{Gessert et~al.(2020)Gessert, Nielsen, Shaikh, Werner and
  Schlaefer}]{gessert2020efficientnets}
\bibinfo{author}{Gessert, N.}, \bibinfo{author}{Nielsen, M.},
  \bibinfo{author}{Shaikh, M.}, \bibinfo{author}{Werner, R.},
  \bibinfo{author}{Schlaefer, A.}, \bibinfo{year}{2020}.
\newblock \bibinfo{title}{Skin lesion classification using ensembles of
  multi-resolution efficientnets with meta data}.
\newblock \bibinfo{journal}{MethodsX} \bibinfo{volume}{7},
  \bibinfo{pages}{100864}.
\newblock \URLprefix
  \url{http://www.sciencedirect.com/science/article/pii/S2215016120300832},
  \DOIprefix\doi{https://doi.org/10.1016/j.mex.2020.100864}.
\bibitem[{Gu et~al.(2024)Gu, Zhang, Wang, Chen, Wang, Ge, Jiao, Ye, Jia and
  Wang}]{gu2024multi}
\bibinfo{author}{Gu, R.}, \bibinfo{author}{Zhang, Y.}, \bibinfo{author}{Wang,
  L.}, \bibinfo{author}{Chen, D.}, \bibinfo{author}{Wang, Y.},
  \bibinfo{author}{Ge, R.}, \bibinfo{author}{Jiao, Z.}, \bibinfo{author}{Ye,
  J.}, \bibinfo{author}{Jia, G.}, \bibinfo{author}{Wang, L.},
  \bibinfo{year}{2024}.
\newblock \bibinfo{title}{Mmy-net: a multimodal network exploiting image and
  patient metadata for simultaneous segmentation and diagnosis}.
\newblock \bibinfo{journal}{Multimedia Systems} \bibinfo{volume}{30}.
\newblock \DOIprefix\doi{10.1007/s00530-024-01260-9}.
\bibitem[{Ha et~al.(2020)Ha, Liu and Liu}]{ha2020identifying}
\bibinfo{author}{Ha, Q.}, \bibinfo{author}{Liu, B.}, \bibinfo{author}{Liu, F.},
  \bibinfo{year}{2020}.
\newblock \bibinfo{title}{Identifying melanoma images using efficientnet
  ensemble: Winning solution to the siim-isic melanoma classification
  challenge}.
\newblock \bibinfo{journal}{arXiv preprint arXiv:2010.05351} .
\bibitem[{Iversen et~al.(2020)Iversen, Igland, Smith-Strøm, Østbye, Tell,
  Skeie, Cooper, Peyrot and Graue}]{iversen2020telemedicine}
\bibinfo{author}{Iversen, M.}, \bibinfo{author}{Igland, J.},
  \bibinfo{author}{Smith-Strøm, H.}, \bibinfo{author}{Østbye, T.},
  \bibinfo{author}{Tell, G.}, \bibinfo{author}{Skeie, S.},
  \bibinfo{author}{Cooper, J.}, \bibinfo{author}{Peyrot, M.},
  \bibinfo{author}{Graue, M.}, \bibinfo{year}{2020}.
\newblock \bibinfo{title}{Effect of a telemedicine intervention for
  diabetes-related foot ulcers on health, well-being and quality of life:
  secondary outcomes from a cluster randomized controlled trial (diafoto)}.
\newblock \bibinfo{journal}{BMC Endocrine Disorders} \bibinfo{volume}{20}.
\newblock \DOIprefix\doi{10.1186/s12902-020-00637-x}.
\bibitem[{Iversen et~al.(2015)Iversen, Tell, Espehaug, Midthjell, Graue, Rokne,
  Berge and Østbye}]{iversen2015depression}
\bibinfo{author}{Iversen, M.M.}, \bibinfo{author}{Tell, G.S.},
  \bibinfo{author}{Espehaug, B.}, \bibinfo{author}{Midthjell, K.},
  \bibinfo{author}{Graue, M.}, \bibinfo{author}{Rokne, B.},
  \bibinfo{author}{Berge, L.I.}, \bibinfo{author}{Østbye, T.},
  \bibinfo{year}{2015}.
\newblock \bibinfo{title}{Is depression a risk factor for diabetic foot ulcers?
  11-years follow-up of the nord-trøndelag health study (hunt)}.
\newblock \bibinfo{journal}{Journal of diabetes and its complications}
  \bibinfo{volume}{29}, \bibinfo{pages}{20—25}.
\newblock \URLprefix \url{https://doi.org/10.1016/j.jdiacomp.2014.09.006},
  \DOIprefix\doi{10.1016/j.jdiacomp.2014.09.006}.
\bibitem[{Jaworek-Korjakowska et~al.(2023)Jaworek-Korjakowska, Wojcicka,
  Kucharski, Brodzicki, Kendrick, Cassidy and Yap}]{jaworek2023hair}
\bibinfo{author}{Jaworek-Korjakowska, J.}, \bibinfo{author}{Wojcicka, A.},
  \bibinfo{author}{Kucharski, D.}, \bibinfo{author}{Brodzicki, A.},
  \bibinfo{author}{Kendrick, C.}, \bibinfo{author}{Cassidy, B.},
  \bibinfo{author}{Yap, M.H.}, \bibinfo{year}{2023}.
\newblock \bibinfo{title}{Skin\_Hair Dataset: Setting the Benchmark for
  Effective Hair Inpainting Methods for Improving the Image Quality of
  Dermoscopic Images}.
\newblock pp. \bibinfo{pages}{167--184}.
\newblock \DOIprefix\doi{10.1007/978-3-031-25069-9_12}.
\bibitem[{Kendrick et~al.(2025)Kendrick, Cassidy, Pappachan, O'Shea, Fernandez,
  Chacko, Jacob, Reeves and Yap}]{kendrick2025segmentation}
\bibinfo{author}{Kendrick, C.}, \bibinfo{author}{Cassidy, B.},
  \bibinfo{author}{Pappachan, J.M.}, \bibinfo{author}{O'Shea, C.},
  \bibinfo{author}{Fernandez, C.J.}, \bibinfo{author}{Chacko, E.},
  \bibinfo{author}{Jacob, K.}, \bibinfo{author}{Reeves, N.D.},
  \bibinfo{author}{Yap, M.H.}, \bibinfo{year}{2025}.
\newblock \bibinfo{title}{Translating clinical delineation of diabetic foot
  ulcers into machine interpretable segmentation}, in: \bibinfo{editor}{Yap,
  M.H.}, \bibinfo{editor}{Kendrick, C.}, \bibinfo{editor}{Br{\"u}ngel, R.}
  (Eds.), \bibinfo{booktitle}{Diabetic Foot Ulcers Grand Challenge},
  \bibinfo{publisher}{Springer Nature Switzerland}, \bibinfo{address}{Cham}.
  pp. \bibinfo{pages}{1--14}.
\bibitem[{Larsson et~al.(1998)Larsson, Agardh, Apelqvist and
  Stenström}]{larsson1998prognosis}
\bibinfo{author}{Larsson, J.}, \bibinfo{author}{Agardh, C.D.},
  \bibinfo{author}{Apelqvist, J.}, \bibinfo{author}{Stenström, A.},
  \bibinfo{year}{1998}.
\newblock \bibinfo{title}{Long term prognosis after healed amputation in
  patients with diabetes}.
\newblock \bibinfo{journal}{Clinical orthopaedics and related research}
  \bibinfo{volume}{350}, \bibinfo{pages}{149--58}.
\newblock \DOIprefix\doi{10.1097/00003086-199805000-00021}.
\bibitem[{Latz et~al.(2019)Latz, Eisenberger and Ullmann}]{latz2019gaussian}
\bibinfo{author}{Latz, J.}, \bibinfo{author}{Eisenberger, M.},
  \bibinfo{author}{Ullmann, E.}, \bibinfo{year}{2019}.
\newblock \bibinfo{title}{Fast sampling of parameterised gaussian random
  fields}.
\newblock \bibinfo{journal}{Computer Methods in Applied Mechanics and
  Engineering} \bibinfo{volume}{348}, \bibinfo{pages}{978--1012}.
\newblock \URLprefix
  \url{https://www.sciencedirect.com/science/article/pii/S004578251930074X},
  \DOIprefix\doi{https://doi.org/10.1016/j.cma.2019.02.003}.
\bibitem[{Le~Ravalec et~al.(2000)Le~Ravalec, Noetinger and
  Hu}]{leravalec2000fft}
\bibinfo{author}{Le~Ravalec, M.}, \bibinfo{author}{Noetinger, B.},
  \bibinfo{author}{Hu, L.}, \bibinfo{year}{2000}.
\newblock \bibinfo{title}{The fft moving average (fft-ma) generator: An
  efficient numerical method for generating and conditioning gaussian
  simulations}.
\newblock \bibinfo{journal}{Mathematical Geology} \bibinfo{volume}{32},
  \bibinfo{pages}{701--723}.
\newblock \DOIprefix\doi{10.1023/A:1007542406333}.
\bibitem[{Lemay et~al.(2021)Lemay, Gros, Vincent, Liu, Cohen and
  Cohen-Adad}]{lemay2021metadata}
\bibinfo{author}{Lemay, A.}, \bibinfo{author}{Gros, C.},
  \bibinfo{author}{Vincent, O.}, \bibinfo{author}{Liu, Y.},
  \bibinfo{author}{Cohen, J.P.}, \bibinfo{author}{Cohen-Adad, J.},
  \bibinfo{year}{2021}.
\newblock \bibinfo{title}{Benefits of linear conditioning for segmentation
  using metadata}, in: \bibinfo{booktitle}{Medical Imaging with Deep Learning}.
\newblock \URLprefix \url{https://openreview.net/forum?id=fa176bQAbr}.
\bibitem[{Li et~al.(2012)Li, Lu, Yang and Wang}]{li2012cloud}
\bibinfo{author}{Li, Q.}, \bibinfo{author}{Lu, W.}, \bibinfo{author}{Yang, J.},
  \bibinfo{author}{Wang, J.Z.}, \bibinfo{year}{2012}.
\newblock \bibinfo{title}{Thin cloud detection of all-sky images using markov
  random fields}.
\newblock \bibinfo{journal}{IEEE Geoscience and Remote Sensing Letters}
  \bibinfo{volume}{9}, \bibinfo{pages}{417--421}.
\newblock \DOIprefix\doi{10.1109/LGRS.2011.2170953}.
\bibitem[{Liu et~al.(2019)Liu, Li, Sun and Yu}]{liu2019gaussian}
\bibinfo{author}{Liu, Y.}, \bibinfo{author}{Li, J.}, \bibinfo{author}{Sun, S.},
  \bibinfo{author}{Yu, B.}, \bibinfo{year}{2019}.
\newblock \bibinfo{title}{Advances in gaussian random field generation: A
  review}.
\newblock \bibinfo{journal}{Computational Geosciences}
  \DOIprefix\doi{10.1007/s10596-019-09867-y}.
\bibitem[{Marzban et~al.(2022)Marzban, Najafi, Agolli and
  Ashrafi}]{marzban2022patient}
\bibinfo{author}{Marzban, S.}, \bibinfo{author}{Najafi, M.},
  \bibinfo{author}{Agolli, A.}, \bibinfo{author}{Ashrafi, E.},
  \bibinfo{year}{2022}.
\newblock \bibinfo{title}{Impact of patient engagement on healthcare quality: A
  scoping review}.
\newblock \bibinfo{journal}{Journal of Patient Experience} \bibinfo{volume}{9},
  \bibinfo{pages}{237437352211254}.
\newblock \DOIprefix\doi{10.1177/23743735221125439}.
\bibitem[{McBride et~al.(2024)McBride, Cassidy, Kendrick, Reeves, Pappachan and
  Yap}]{mcbride2024colour}
\bibinfo{author}{McBride, C.}, \bibinfo{author}{Cassidy, B.},
  \bibinfo{author}{Kendrick, C.}, \bibinfo{author}{Reeves, N.D.},
  \bibinfo{author}{Pappachan, J.M.}, \bibinfo{author}{Yap, M.H.},
  \bibinfo{year}{2024}.
\newblock \bibinfo{title}{Multi-colour space channel selection for improved
  chronic wound segmentation}, in: \bibinfo{booktitle}{2024 IEEE International
  Symposium on Biomedical Imaging (ISBI)}, pp. \bibinfo{pages}{1--5}.
\newblock \DOIprefix\doi{10.1109/ISBI56570.2024.10635155}.
\bibitem[{{Ministry of Housing, Communities \& Local
  Government}(2019)}]{2019deprivation}
\bibinfo{author}{{Ministry of Housing, Communities \& Local Government}},
  \bibinfo{year}{2019}.
\newblock \bibinfo{title}{English indices of deprivation 2019}.
\newblock \URLprefix
  \url{https://imd-by-postcode.opendatacommunities.org/imd/2019}.
  \bibinfo{note}{last access: 02/12/24}.
\bibitem[{Moura et~al.(2019)Moura, Rodrigues, Gonçalves, Amaral, Lima and
  Carvalho}]{moura2019dfu}
\bibinfo{author}{Moura, J.}, \bibinfo{author}{Rodrigues, J.},
  \bibinfo{author}{Gonçalves, M.}, \bibinfo{author}{Amaral, C.},
  \bibinfo{author}{Lima, M.}, \bibinfo{author}{Carvalho, E.},
  \bibinfo{year}{2019}.
\newblock \bibinfo{title}{Imbalance in t-cell differentiation as a biomarker of
  chronic diabetic foot ulceration}.
\newblock \bibinfo{journal}{Cellular \& Molecular Immunology} ,
  \bibinfo{pages}{1--2}.
\bibitem[{Nur Anggraini~Ningrum et~al.(2021)Nur Anggraini~Ningrum, Yuan, Kung,
  Wu, Tzeng, Huang, Li and Wang}]{ningrum2021metadata}
\bibinfo{author}{Nur Anggraini~Ningrum, D.}, \bibinfo{author}{Yuan, S.P.},
  \bibinfo{author}{Kung, W.M.}, \bibinfo{author}{Wu, C.C.},
  \bibinfo{author}{Tzeng, I.S.}, \bibinfo{author}{Huang, C.Y.},
  \bibinfo{author}{Li, Y.C.}, \bibinfo{author}{Wang, Y.C.R.},
  \bibinfo{year}{2021}.
\newblock \bibinfo{title}{Deep learning classifier with patient’s metadata of
  dermoscopic images in malignant melanoma detection}.
\newblock \bibinfo{journal}{Journal of Multidisciplinary Healthcare}
  \bibinfo{volume}{2021}, \bibinfo{pages}{877--885}.
\newblock \DOIprefix\doi{10.2147/JMDH.S306284}.
\bibitem[{Okafor et~al.(2024)Okafor, Cassidy, O'Shea and
  Pappachan}]{okafor2024preproc}
\bibinfo{author}{Okafor, N.C.}, \bibinfo{author}{Cassidy, B.},
  \bibinfo{author}{O'Shea, C.}, \bibinfo{author}{Pappachan, J.M.},
  \bibinfo{year}{2024}.
\newblock \bibinfo{title}{The Effect of Image Preprocessing Algorithms on
  Diabetic Foot Ulcer Classification}.
\newblock pp. \bibinfo{pages}{336--352}.
\newblock \DOIprefix\doi{10.1007/978-3-031-66958-3_25}.
\bibitem[{Pappachan et~al.(2022)Pappachan, Cassidy, Fernandez, Chandrabalan and
  Yap}]{joseph2022future}
\bibinfo{author}{Pappachan, J.M.}, \bibinfo{author}{Cassidy, B.},
  \bibinfo{author}{Fernandez, C.J.}, \bibinfo{author}{Chandrabalan, V.},
  \bibinfo{author}{Yap, M.H.}, \bibinfo{year}{2022}.
\newblock \bibinfo{title}{The role of artificial intelligence technology in the
  care of diabetic foot ulcers: the past, the present, and the future}.
\newblock \bibinfo{journal}{World Journal of Diabetes} \bibinfo{volume}{13},
  \bibinfo{pages}{1131--1139}.
\newblock \DOIprefix\doi{10.4239/wjd.v13.i12.1131}.
\bibitem[{Patel et~al.(2024)Patel, Shah, Dhar, Zhang, Niezgoda, Gopalakrishnan
  and Yu}]{patel2024wound}
\bibinfo{author}{Patel, Y.}, \bibinfo{author}{Shah, T.}, \bibinfo{author}{Dhar,
  M.}, \bibinfo{author}{Zhang, T.}, \bibinfo{author}{Niezgoda, J.},
  \bibinfo{author}{Gopalakrishnan, S.}, \bibinfo{author}{Yu, Z.},
  \bibinfo{year}{2024}.
\newblock \bibinfo{title}{Integrated image and location analysis for wound
  classification: a deep learning approach}.
\newblock \bibinfo{journal}{Scientific Reports} \bibinfo{volume}{14}.
\newblock \DOIprefix\doi{10.1038/s41598-024-56626-w}.
\bibitem[{Petersen et~al.(2022)Petersen, Linde-Zwirble, Tan, Rothenberg,
  Salgado, Bloom and Armstrong}]{petersen2022mortality}
\bibinfo{author}{Petersen, B.}, \bibinfo{author}{Linde-Zwirble, W.},
  \bibinfo{author}{Tan, T.W.}, \bibinfo{author}{Rothenberg, G.},
  \bibinfo{author}{Salgado, S.}, \bibinfo{author}{Bloom, J.},
  \bibinfo{author}{Armstrong, D.}, \bibinfo{year}{2022}.
\newblock \bibinfo{title}{Higher rates of all-cause mortality and resource
  utilization during episodes-of-care for diabetic foot ulceration}.
\newblock \bibinfo{journal}{Diabetes Research and Clinical Practice}
  \DOIprefix\doi{10.1016/j.diabres.2021.109182}.
\bibitem[{Petrone et~al.(2021)Petrone, Giribono, Massini, Pietrangelo,
  Magnifico, Bracale, Di~Marco, Bracale and
  Petronio~Petronio}]{petrone2021foot}
\bibinfo{author}{Petrone, F.}, \bibinfo{author}{Giribono, A.},
  \bibinfo{author}{Massini, L.}, \bibinfo{author}{Pietrangelo, L.},
  \bibinfo{author}{Magnifico, I.}, \bibinfo{author}{Bracale, U.},
  \bibinfo{author}{Di~Marco, R.}, \bibinfo{author}{Bracale, R.},
  \bibinfo{author}{Petronio~Petronio, G.}, \bibinfo{year}{2021}.
\newblock \bibinfo{title}{Retrospective observational study on microbial
  contamination of ulcerative foot lesions in diabetic patients}.
\newblock \bibinfo{journal}{Microbiology Research} \bibinfo{volume}{12}.
\newblock \DOIprefix\doi{10.3390/microbiolres12040058}.
\bibitem[{Pewton et~al.(2024)Pewton, Cassidy, Kendrick and Yap}]{pewton2024dca}
\bibinfo{author}{Pewton, S.W.}, \bibinfo{author}{Cassidy, B.},
  \bibinfo{author}{Kendrick, C.}, \bibinfo{author}{Yap, M.H.},
  \bibinfo{year}{2024}.
\newblock \bibinfo{title}{Dermoscopic dark corner artifacts removal: Friend or
  foe?}
\newblock \bibinfo{journal}{Computer Methods and Programs in Biomedicine}
  \bibinfo{volume}{244}, \bibinfo{pages}{107986}.
\newblock \URLprefix
  \url{https://www.sciencedirect.com/science/article/pii/S0169260723006521},
  \DOIprefix\doi{https://doi.org/10.1016/j.cmpb.2023.107986}.
\bibitem[{Polikandrioti et~al.(2020)Polikandrioti, Vasilopoulos, Koutelekos,
  Panoutsopoulos, Gerogianni, Alikari, Dousis and
  Zartaloudi}]{polikandrioti2020depression}
\bibinfo{author}{Polikandrioti, M.}, \bibinfo{author}{Vasilopoulos, G.},
  \bibinfo{author}{Koutelekos, I.}, \bibinfo{author}{Panoutsopoulos, G.},
  \bibinfo{author}{Gerogianni, G.}, \bibinfo{author}{Alikari, V.},
  \bibinfo{author}{Dousis, E.}, \bibinfo{author}{Zartaloudi, A.},
  \bibinfo{year}{2020}.
\newblock \bibinfo{title}{Depression in diabetic foot ulcer: Associated factors
  and the impact of perceived social support and anxiety on depression}.
\newblock \bibinfo{journal}{International Wound Journal} \bibinfo{volume}{17},
  \bibinfo{pages}{900--909}.
\newblock \URLprefix
  \url{https://onlinelibrary.wiley.com/doi/abs/10.1111/iwj.13348},
  \DOIprefix\doi{https://doi.org/10.1111/iwj.13348}.
\bibitem[{Rathur and Boulton(2007)}]{rathur2007neuropathic}
\bibinfo{author}{Rathur, H.}, \bibinfo{author}{Boulton, A.},
  \bibinfo{year}{2007}.
\newblock \bibinfo{title}{The neuropathic diabetic foot}.
\newblock \bibinfo{journal}{Nature clinical practice. Endocrinology \&
  metabolism} \bibinfo{volume}{3}, \bibinfo{pages}{14--25}.
\bibitem[{Reeves et~al.(2021)Reeves, Cassidy, Abbott and
  Yap}]{reeves2021diabetes}
\bibinfo{author}{Reeves, N.D.}, \bibinfo{author}{Cassidy, B.},
  \bibinfo{author}{Abbott, C.A.}, \bibinfo{author}{Yap, M.H.},
  \bibinfo{year}{2021}.
\newblock \bibinfo{title}{Chapter 7 - novel technologies for detection and
  prevention of diabetic foot ulcers}, in: \bibinfo{editor}{Gefen, A.} (Ed.),
  \bibinfo{booktitle}{The Science, Etiology and Mechanobiology of Diabetes and
  its Complications}. \bibinfo{publisher}{Academic Press}, pp.
  \bibinfo{pages}{107--122}.
\newblock \URLprefix
  \url{https://www.sciencedirect.com/science/article/pii/B9780128210703000076},
  \DOIprefix\doi{https://doi.org/10.1016/B978-0-12-821070-3.00007-6}.
\bibitem[{Renner and Erfurt-Berge(2017)}]{renner2017depression}
\bibinfo{author}{Renner, R.}, \bibinfo{author}{Erfurt-Berge, C.},
  \bibinfo{year}{2017}.
\newblock \bibinfo{title}{Depression and quality of life in patients with
  chronic wounds: ways to measure their influence and their effect on daily
  life}.
\newblock \bibinfo{journal}{Chronic Wound Care Management and Research}
  \bibinfo{volume}{4}, \bibinfo{pages}{143--151}.
\newblock \DOIprefix\doi{https://doi.org/10.2147/CWCMR.S124917}.
\bibitem[{Sen(2021)}]{sen2021wound}
\bibinfo{author}{Sen, C.K.}, \bibinfo{year}{2021}.
\newblock \bibinfo{title}{Human wound and its burden: Updated 2020 compendium
  of estimates}.
\newblock \bibinfo{journal}{Advances in Wound Care} \bibinfo{volume}{10},
  \bibinfo{pages}{281--292}.
\newblock \URLprefix \url{https://doi.org/10.1089/wound.2021.0026},
  \DOIprefix\doi{10.1089/wound.2021.0026}. \bibinfo{note}{pMID: 33733885}.
\bibitem[{Serban et~al.(2020)Serban, Papanas, Dascalu, Stana, Vanessa, Vancea,
  Badiu, Tanasescu, Tudor, Balasescu and Pantea~Stoian}]{serban2020retinopahy}
\bibinfo{author}{Serban, D.}, \bibinfo{author}{Papanas, N.},
  \bibinfo{author}{Dascalu, A.}, \bibinfo{author}{Stana, D.},
  \bibinfo{author}{Vanessa, N.}, \bibinfo{author}{Vancea, G.},
  \bibinfo{author}{Badiu, C.}, \bibinfo{author}{Tanasescu, D.},
  \bibinfo{author}{Tudor, C.}, \bibinfo{author}{Balasescu, S.},
  \bibinfo{author}{Pantea~Stoian, A.}, \bibinfo{year}{2020}.
\newblock \bibinfo{title}{Diabetic retinopathy in patients with diabetic foot
  ulcer: A systematic review}.
\newblock \bibinfo{journal}{The International Journal of Lower Extremity
  Wounds} \bibinfo{volume}{20}, \bibinfo{pages}{153473462098223}.
\newblock \DOIprefix\doi{10.1177/1534734620982237}.
\bibitem[{Sharma et~al.(2019)Sharma, Vans, Shigemizu, Boroevich and
  Tsunoda}]{sharma2019deepinsight}
\bibinfo{author}{Sharma, A.}, \bibinfo{author}{Vans, E.},
  \bibinfo{author}{Shigemizu, D.}, \bibinfo{author}{Boroevich, K.},
  \bibinfo{author}{Tsunoda, T.}, \bibinfo{year}{2019}.
\newblock \bibinfo{title}{Deepinsight: A methodology to transform a non-image
  data to an image for convolution neural network architecture}.
\newblock \bibinfo{journal}{Scientific Reports} \bibinfo{volume}{9}.
\newblock \DOIprefix\doi{10.1038/s41598-019-47765-6}.
\bibitem[{Sheffield(2004)}]{scott2004fields}
\bibinfo{author}{Sheffield, S.}, \bibinfo{year}{2004}.
\newblock \bibinfo{title}{Gaussian free fields for mathematicians}.
\newblock \bibinfo{journal}{Probability Theory and Related Fields}
  \bibinfo{volume}{139}.
\newblock \DOIprefix\doi{10.1007/s00440-006-0050-1}.
\bibitem[{Stefanou(2009)}]{stefanou2009stochastic}
\bibinfo{author}{Stefanou, G.}, \bibinfo{year}{2009}.
\newblock \bibinfo{title}{The stochastic finite element method: Past, present
  and future}.
\newblock \bibinfo{journal}{Computer Methods in Applied Mechanics and
  Engineering} \bibinfo{volume}{198}, \bibinfo{pages}{1031--1051}.
\newblock \URLprefix
  \url{https://www.sciencedirect.com/science/article/pii/S0045782508004118},
  \DOIprefix\doi{https://doi.org/10.1016/j.cma.2008.11.007}.
\bibitem[{Su et~al.(2018)Su, Chang, Yunshing and Chen}]{su2018dfu}
\bibinfo{author}{Su, C.L.}, \bibinfo{author}{Chang, C.C.},
  \bibinfo{author}{Yunshing, P.}, \bibinfo{author}{Chen, M.Y.},
  \bibinfo{year}{2018}.
\newblock \bibinfo{title}{The predictive factors associated with comorbidities
  for treatment response in outpatients with king classification iii diabetes
  foot ulcers}.
\newblock \bibinfo{journal}{Annals of Plastic Surgery} \bibinfo{volume}{81},
  \bibinfo{pages}{S39--S43}.
\newblock \DOIprefix\doi{10.1097/SAP.0000000000001500}.
\bibitem[{Tang et~al.(2015)Tang, Li, Zhou and Phoon}]{tang2015slope}
\bibinfo{author}{Tang, X.S.}, \bibinfo{author}{Li, D.Q.},
  \bibinfo{author}{Zhou, C.B.}, \bibinfo{author}{Phoon, K.K.},
  \bibinfo{year}{2015}.
\newblock \bibinfo{title}{Copula-based approaches for evaluating slope
  reliability under incomplete probability information}.
\newblock \bibinfo{journal}{Structural Safety} \bibinfo{volume}{52},
  \bibinfo{pages}{90--99}.
\newblock \URLprefix
  \url{https://www.sciencedirect.com/science/article/pii/S0167473014000903},
  \DOIprefix\doi{https://doi.org/10.1016/j.strusafe.2014.09.007}.
\bibitem[{Vainieri et~al.(2020)Vainieri, Ahluwalia, Slim, Walton, Manu, Taori,
  Wilkins, Huang, Edmonds, Rashid, Kavarthapu and Vas}]{vainieri2020mortality}
\bibinfo{author}{Vainieri, E.}, \bibinfo{author}{Ahluwalia, R.},
  \bibinfo{author}{Slim, H.}, \bibinfo{author}{Walton, D.},
  \bibinfo{author}{Manu, C.}, \bibinfo{author}{Taori, S.},
  \bibinfo{author}{Wilkins, J.}, \bibinfo{author}{Huang, D.},
  \bibinfo{author}{Edmonds, M.}, \bibinfo{author}{Rashid, H.},
  \bibinfo{author}{Kavarthapu, V.}, \bibinfo{author}{Vas, P.},
  \bibinfo{year}{2020}.
\newblock \bibinfo{title}{Outcomes after emergency admission with a diabetic
  foot attack indicate a high rate of healing and limb salvage but increased
  mortality: 18-month follow-up study}.
\newblock \bibinfo{journal}{Experimental and Clinical Endocrinology \&
  Diabetes} \DOIprefix\doi{10.1055/a-1322-4811}.
\bibitem[{Van~Netten et~al.(2017)Van~Netten, Clark, Lazzarini, Janda and
  Reed}]{netten2017mobile}
\bibinfo{author}{Van~Netten, J.}, \bibinfo{author}{Clark, D.},
  \bibinfo{author}{Lazzarini, P.}, \bibinfo{author}{Janda, M.},
  \bibinfo{author}{Reed, L.}, \bibinfo{year}{2017}.
\newblock \bibinfo{title}{The validity and reliability of remote diabetic foot
  ulcer assessment using mobile phone images}.
\newblock \bibinfo{journal}{Scientific Reports} \bibinfo{volume}{7}.
\newblock \DOIprefix\doi{10.1038/s41598-017-09828-4}.
\bibitem[{Xi et~al.(2012)Xi, Jung and Youn}]{xi2012random}
\bibinfo{author}{Xi, Z.}, \bibinfo{author}{Jung, B.C.}, \bibinfo{author}{Youn,
  B.D.}, \bibinfo{year}{2012}.
\newblock \bibinfo{title}{Random field modeling with insufficient data sets for
  probability analysis}, in: \bibinfo{booktitle}{2012 Proceedings Annual
  Reliability and Maintainability Symposium}, pp. \bibinfo{pages}{1--5}.
\newblock \DOIprefix\doi{10.1109/RAMS.2012.6175482}.
\bibitem[{Yammine and Estephan(2021)}]{yammine2021telemedicine}
\bibinfo{author}{Yammine, K.}, \bibinfo{author}{Estephan, M.},
  \bibinfo{year}{2021}.
\newblock \bibinfo{title}{Telemedicine and diabetic foot ulcer outcomes. a
  meta-analysis of controlled trials}.
\newblock \bibinfo{journal}{The Foot} \URLprefix
  \url{https://www.sciencedirect.com/science/article/pii/S0958259221000985},
  \DOIprefix\doi{https://doi.org/10.1016/j.foot.2021.101872}.
\bibitem[{Yang et~al.(2023)Yang, Yang, Zhang and Wang}]{yang2023markov}
\bibinfo{author}{Yang, X.}, \bibinfo{author}{Yang, X.}, \bibinfo{author}{Zhang,
  C.}, \bibinfo{author}{Wang, J.}, \bibinfo{year}{2023}.
\newblock \bibinfo{title}{Sar image classification using markov random fields
  with deep learning}.
\newblock \bibinfo{journal}{Remote Sensing} \bibinfo{volume}{15}.
\newblock \URLprefix \url{https://www.mdpi.com/2072-4292/15/3/617},
  \DOIprefix\doi{10.3390/rs15030617}.
\bibitem[{Yap et~al.(2021)Yap, Hachiuma, Alavi, Brüngel, Cassidy, Goyal, Zhu,
  Rückert, Olshansky, Huang, Saito, Hassanpour, Friedrich, Ascher, Song,
  Kajita, Gillespie, Reeves, Pappachan, O'Shea and Frank}]{yap2021evaluation}
\bibinfo{author}{Yap, M.H.}, \bibinfo{author}{Hachiuma, R.},
  \bibinfo{author}{Alavi, A.}, \bibinfo{author}{Brüngel, R.},
  \bibinfo{author}{Cassidy, B.}, \bibinfo{author}{Goyal, M.},
  \bibinfo{author}{Zhu, H.}, \bibinfo{author}{Rückert, J.},
  \bibinfo{author}{Olshansky, M.}, \bibinfo{author}{Huang, X.},
  \bibinfo{author}{Saito, H.}, \bibinfo{author}{Hassanpour, S.},
  \bibinfo{author}{Friedrich, C.M.}, \bibinfo{author}{Ascher, D.B.},
  \bibinfo{author}{Song, A.}, \bibinfo{author}{Kajita, H.},
  \bibinfo{author}{Gillespie, D.}, \bibinfo{author}{Reeves, N.D.},
  \bibinfo{author}{Pappachan, J.M.}, \bibinfo{author}{O'Shea, C.},
  \bibinfo{author}{Frank, E.}, \bibinfo{year}{2021}.
\newblock \bibinfo{title}{Deep learning in diabetic foot ulcers detection: A
  comprehensive evaluation}.
\newblock \bibinfo{journal}{Computers in Biology and Medicine}
  \bibinfo{volume}{135}, \bibinfo{pages}{104596}.
\newblock \URLprefix
  \url{https://www.sciencedirect.com/science/article/pii/S0010482521003905},
  \DOIprefix\doi{https://doi.org/10.1016/j.compbiomed.2021.104596}.
\bibitem[{Zhu et~al.(2003)Zhu, Lafferty and Ghahramani}]{zhu2003gaussian}
\bibinfo{author}{Zhu, X.}, \bibinfo{author}{Lafferty, J.},
  \bibinfo{author}{Ghahramani, Z.}, \bibinfo{year}{2003}.
\newblock \bibinfo{title}{Combining active learning and semi-supervised
  learning using gaussian fields and harmonic functions} .

\end{thebibliography}



\end{document}